\documentclass{cernyrep}
\usepackage[T1]{fontenc}
\usepackage[bookmarks, colorlinks=true, linktoc=page, pdftex, linkcolor=black, citecolor=black, urlcolor=blue]{hyperref}

\usepackage{fancyhdr}
\usepackage{empheq}
\usepackage{subfigure,units}
\fancyhfoffset{4 mm}
\fancypagestyle{ARTTITLE}{%
\fancyhf{} 
\lhead{\hfill CERN Accelerator School Proceedings ---
{\it RF for Accelerators} ---  Berlin,  Germany, 2023\hfill}
\lfoot{\hspace{3mm} Available online at \url{https://cas.web.cern.ch/previous-schools}}
\rfoot{\thepage\hspace*{3mm}}
 
}

\usepackage{varwidth}
\usepackage{xcolor}
\newcommand{\dwt}{\frac{\mathrm{d}}{\mathrm{d}t}}

\begin{document}

\title{Electromagnetic Simulations}
\author{Thomas Flisgen}
\institute{Brandenburg University of Technology Cottbus-Senftenberg\\Ferdinand-Braun-Institut gGmbH}

\begin{abstract}
This contribution reviews fundamental concepts of electromagnetic simulations for RF structures used in particle accelerators. The necessity of numerical methods for electromagnetic simulations is discussed, and the fundamental steps involved in performing such simulations are presented. Spatial discretization is reviewed in a very general way to obtain semi-discrete equations. These semi-discrete equations are employed to revisit time-domain and frequency-domain approaches. Special attention is devoted to the computation of network matrices (e.g.\ scattering matrices) of structures with large quality factors as this topic is highly relevant in the context of radio-frequency for accelerators. This leads to a very short recapitulation of model-order reduction methods. The various sources of error inherent in electromagnetic simulations are examined. Finally, recommendations for the careful validation and assessment of simulation results are presented.
\end{abstract}

\maketitle
\thispagestyle{ARTTITLE}
\section{Introduction}
Electromagnetic simulations are an essential tool for the design of radio-frequency (RF) systems for particle accelerators, because they allow to study the behavior of electromagnetic fields inside complex structures. Electromagnetic simulations require numerical solutions of Maxwell's equations under given boundary and excitation conditions. Computational Electromagnetics or Computational Electromagnetism (CEM) is the field of research dealing with numerical methods to solve Maxwell's equations. It has been a very active field of research for more than fifty years. CEM is not only driven by accelerator physics, but has also been advanced by the development of antennas, radar systems, mobile phones, and many other modern devices using electromagnetic fields. Foundations of CEM can be found in~Refs.~\cite{rylander2012computational,sheng2012essentials,Davidson2005} and valuable overview papers on electromagnetic simulations are provided in Refs.~\cite{Sumithra2017CEMReview,Weiland2017,Munteanu2010,Weiland2008}. Over the past decades, powerful computing hardware has become available and efficient numerical methods have matured. Consequently, there are many different methods for numerically solving Maxwell's equations, such as Finite-Difference Time-Domain (FDTD) method~\cite{yee1966numerical,taflove2000computational}, Finite Integration Technique (FIT)~\cite{SchuhmannWeiland2004,weiland1996timedomain,weiland1977discretization}, Finite Element Method (FEM)~\cite{Jin,volakis1998electromagnetics,polycarpou2006introduction} and Method of Moments (MoM)~\cite{gibson2008method,chew2001fast}, as well as modifications of these methods. Widely used commercial software packages, such as CST Studio Suite\textregistered~\cite{CST}, HFSS~\cite{HFSS}, COMSOL Multiphysics\textregistered~\cite{COMSOL}, EMPIRE XPU~\cite{EMPIRE}, and EMPro~\cite{EMPro}, provide efficient and validated implementations of selected numerical methods for real-world electromagnetic simulations. Complementary to commercial software packages, open-source environments, such as openEMS~\cite{openEMS}, openCFS~\cite{openCFS}, Meep~\cite{Meep}, FeNiCs~\cite{FEniCS}, and Netgen/NGSolve~\cite{NGSolve}, allow for electromagnetic simulations.

A detailed introduction to the aforementioned methods and to the software packages, including their respective advantages and disadvantages, is beyond the scope of this contribution, as it cannot be covered comprehensively. Instead, the topic is explored in a general fashion for volume-based methods, such as FDTD, FIT, and FEM, because these approaches are especially relevant in the domain of RF for accelerators. The contribution is organized as follows: Section~\ref{sec:figofmerit} highlights the importance of solving Maxwell's equations for RF applications in particle accelerators. Section~\ref{sec:analytical} discusses the value of analytical solutions. Section~\ref{sec:workflow} outlines the key steps involved in performing electromagnetic simulations. Section~\ref{sec:semidiscrete} illustrates the process of transforming the partial differential equations governing the electromagnetic field problem into semi-discrete forms, where the spatial derivatives are approximated and time derivatives remain. Section~\ref{sec:time_do} demonstrates how the semi-discrete formulation can be applied to handle transient electromagnetic fields, whereas Section~\ref{sec:freq_do} shows how electromagnetic fields in frequency domain are treated. Section~\ref{sec:highQ} focuses on simulating highly resonant structures as this is a~crucial task in the context of RF for accelerators. Section~\ref{sec:errors} addresses the errors arising in electromagnetic simulations, whereas Section~\ref{sec:remarkspractice} provides general remarks on the practice of electromagnetic simulations. Finally, Section~\ref{sec:conclusions} presents conclusions. 

\subsection{Figures of Merit of RF Devices in Particle Accelerators\label{sec:figofmerit}}
Particle accelerators are powerful instruments in modern science and technology that are successfully employed in industry, medicine, and in fundamental research. The heart of these machines are their vacuum chambers with their insertion devices. In these chambers, bunches of charged particles are accelerated and deflected by means of electromagnetic fields. The charged particle beam is often accelerated using hollow resonators, also known as cavities. These cavities are typically equipped with power couplers and higher-order mode couplers. The cavity is driven by an RF power source through the input coupler to excite a dedicated resonance. This resonance transfers the energy from the RF source to the~beam of charged particles. The higher-order mode couplers are designed to damp all other (higher-order) resonances in the cavity that can be excited by the beam of charged particles. Some higher-order mode couplers are equipped with notch filters to reject the fundamental accelerating mode while damping higher-order modes.

\begin{figure}
  \centering
  \subfigure[]{\includegraphics[width=0.31\textwidth]{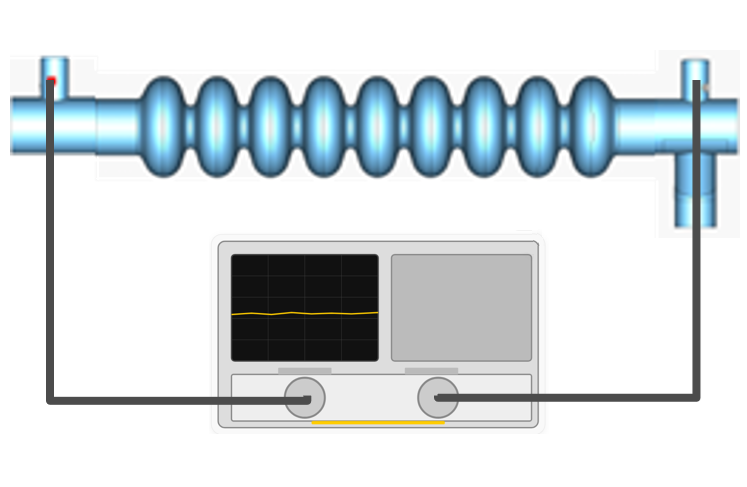}\label{fig:network}}\quad
  \subfigure[]{\includegraphics[width=0.31\textwidth]{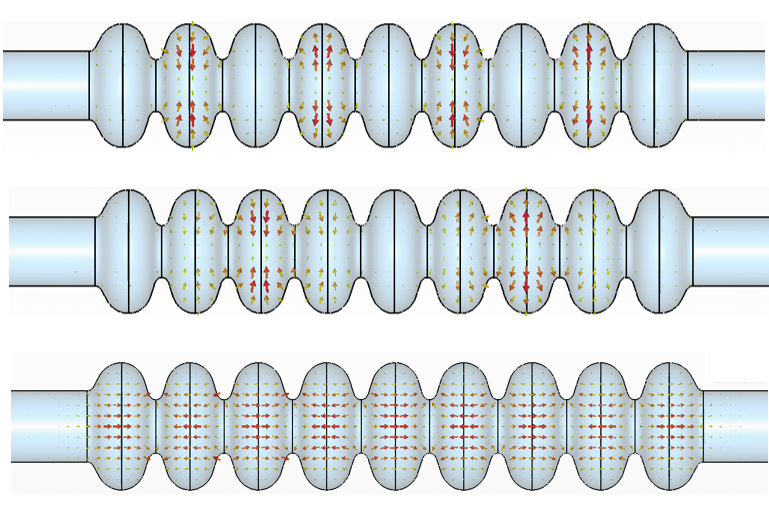}\label{fig:eigenmodes}}\quad
  \subfigure[]{\includegraphics[width=0.31\textwidth]{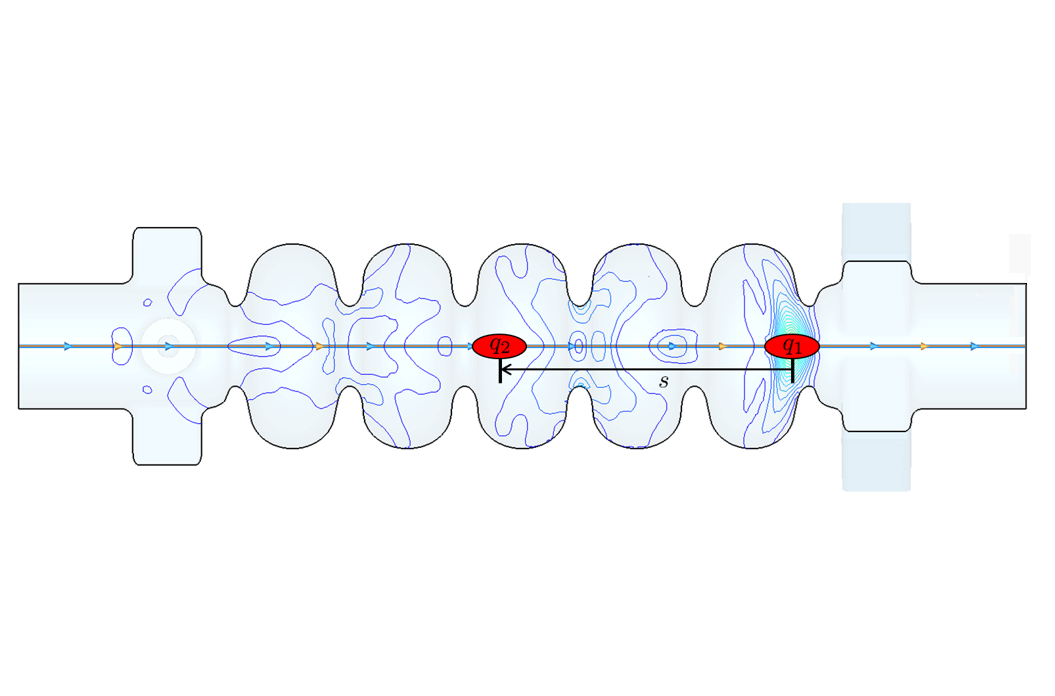}\label{fig:wake}}
  \caption{(a) Cavity with higher-order mode couplers connected to a vector network analyzer. (b) Cavity with field distributions of three example eigenmodes. (c) Wakefields in a cavity (electric field strength isolines) excited by a~leading bunch with charge $q_1$, followed by a trailing bunch with charge $q_2$ at a distance $s$.}
  \label{fig:figuresofmerit}
\end{figure}

It is safe to say that the aforementioned RF systems are complex components. Thus, formalisms for suitably describing their properties are required. As shown in Fig.~\ref{fig:figuresofmerit}, these systems can be characterized using various figures of merit, such as scattering, impedance, and admittance parameters, characteristic impedances arising from frequencies and field distributions of eigenmodes, wake potentials, and wake impedances. These quantities are all defined by integrals over electric and/or magnetic field distributions. The field distributions ultimately result from solving Maxwell's equations (or equations derived from them, such as the curl-curl equation or the wave equation) for the respective boundary and excitation conditions. Maxwell’s equations in differential form, together with the material relations, constitute a coupled system of partial differential equations. Solving these equations is required to analyze and optimize electromagnetic devices and systems according to the appropriate figures of merit.

\subsection{Remarks on Analytical Solutions of Maxwell's Equations\label{sec:analytical}}
Unfortunately, analytical solutions for Maxwell’s equations (as well as for curl-curl equation and wave equation) are only available for very simple symmetric geometries. Analytical field solutions can typically be found by the separation of variables, if boundaries of the geometry coincide with coordinate planes of a suitable coordinate system. As discussed in~Ref.~\cite{ZhangLi2008}, eleven orthogonal coordinate systems are known that allow for a separation of variables. The Cartesian, cylindrical, and spherical coordinate systems are clearly the most prominent among these. Analytical considerations are very important to gain fundamental understanding of field solutions in general: Conservation and scaling laws can be analytically derived and propagation in free space or waveguides can be studied. The theory of electromagnetic fields and of respective analytical solutions can be found in CAS proceedings, such as Refs.~\cite{shreyber2024emtheory,flisgen2018recap,wolski2011emfields,miles2000rf}, or in standard textbooks, such as~Refs.~\cite{Griffiths,Jackson,Collin2,Collin1,Balanis}. A solid understanding of electromagnetic theory and analytical methods for solving Maxwell’s equations is a prerequisite for conducting meaningful electromagnetic simulations. If the geometry of the device under study is so simple, that an analytical solution is available, it should be used! The analytical solution is continuous, so its derivatives are available for optimization. Parameter studies can be performed easily and there are no errors from numerical approximations and almost no computational costs.

\section{Foundations of Electromagnetic Simulations}
Determining field distributions in realistic geometries, such as depicted in Fig.~\ref{fig:figuresofmerit}, generally requires the~application of numerical methods in electromagnetic simulations. The advantage of numerical methods is their flexibility, which allows them to determine field distributions in complex geometries with inhomogeneous material distributions and sophisticated excitation conditions.

\subsection{General Workflow\label{sec:workflow}}
The workflow for conducting electromagnetic simulations is listed hereinafter. The listed steps are of general nature and do not depend on the actual software or even method used for simulation. For clarity, they are illustrated using the analysis of a cavity as an actual example. 

\begin{enumerate}
	\item \textbf{Initial Problem:} A structure is given whose electromagnetic figures of merit are to be determined. In the context of RF accelerator applications, the quantities presented in Section~\ref{sec:figofmerit} may be of interest, among other relevant parameters.
	\item \textbf{Idealization and Simplification:} The structure is idealized and simplified as much as possible. It is essential to ensure that the introduced approximations are justified, which requires some experience. For example, it is often reasonable to assume that the cavity is rotationally symmetric or that the surface losses of a superconducting cavity are zero. The creation of a vacuum model of the~structure is often sufficient, which is a significant simplification of the geometry.
	\item \textbf{Specifying (Partial) Differential Equations:} The governing partial differential equations that describe the field distributions have to be specified. Typically, Maxwell's equations are not directly solved in numerical simulations, but partial differential equations arising from those.  This results in the electro- and magnetostatic case in Poisson equations and in the magnetoquasistatic case in diffusion equations. If the expected wavelength is in the order of the size of the structure or even smaller, as it is the case for accelerating cavities, full wave analysis is required and curl-curl equations or wave equations have to be solved. Boundary and excitation conditions have to be formulated in addition to the partial differential equations. Typical boundary and symmetry conditions in the context of cavities are perfect electric conducting (PEC) and perfect magnetic conducting (PMC) walls. PEC boundaries require tangential electric field components and normal magnetic field components to be zero at the boundary, whereas PMC boundaries require normal electric field components and tangential magnetic field components to be zero. Cavities are typically excited via waveguide ports, where the electromagnetic field at the boundary is prescribed in terms of known waveguide eigenmodes, obtained from the solution of a two-dimensional cross-sectional eigenvalue problem.
	\item \textbf{Discretization:} The governing partial differential equations describe fields, which are continuous in space and time. These partial differential equations and their associated boundary conditions are discretized and represented as a system of matrix–vector equations. Matrix–vector equations can be efficiently handled and solved by digital computers. Section~\ref{sec:semidiscrete} discusses the discretization in greater detail, as it is fundamental to electromagnetic simulations.

Since the accuracy of the numerical solution depends directly on the chosen discretization, mesh convergence and refinement studies are critical. An insufficiently resolved mesh may lead to significant discretization errors, inaccurate field distributions, or incorrect resonance frequencies. Therefore, systematic mesh refinement is required to ensure that the solution converges toward a stable result that is independent of the discretization (see Step 7 and Section~\ref{sec:validation}).
	\item \textbf{Determination of Field Distributions and Derived Quantities:} The solution of the matrix–vector equations (refer to Section~\ref{sec:time_do} and Section~\ref{sec:freq_do}) yields vectors representing the field distributions. These field distributions provide important insight into the electromagnetic behavior of the structures. In addition, scalar quantities (see Section~\ref{sec:figofmerit}) resulting from integrals over the field distributions need to be determined. This step is referred to as post-processing. Modern electromagnetic simulation packages natively include various post-processing options for acquiring derived quantities.
	 \item \textbf{Visualization of Results:} The large numerical datasets generated must be presented in a clear and meaningful manner. Field distributions in the three-dimensional domain are typically visualized on selected cut planes as vector plots or contour plots. Scalar quantities derived from the field distributions are often presented using line plots to illustrate their dependence on parameters such as frequency.
	 \item \textbf{Validation and Interpretation}: The steps described above introduce errors, which are discussed in Section~\ref{sec:errors}. Thus, it is essential to validate numerical results in order to ensure the reliability of the simulation. Given its importance, validation is treated separately in Section~\ref{sec:validation}. Careful interpretation of the results is necessary to gain physical insight into the behavior of the investigated structure and to identify the underlying electromagnetic mechanisms.
\end{enumerate}

\subsection{Semi-discrete Equations\label{sec:semidiscrete}}
This section is about converting partial differential equations into semi-discrete equations. In these equations, the partial derivatives with respect to the spatial coordinates are approximated by means of matrix-vector products, while the derivatives with respect to time are still present. As wave phenomena are relevant in the context of RF accelerators, none of the time-derivatives are set to zero as they would be in static or quasi-static regimes \cite{SteinmetzKurzClemens2011}.

Considering the induction law in its differential representation and replacing the magnetic flux density $\mathbf{B}(\mathbf{r},t)$ by the magnetic field strength $\mathbf{H}(\mathbf{r},t)$ using the material equation  
\begin{equation}
\mu\mathbf{H}(\mathbf{r},t) =\mathbf{B}(\mathbf{r},t)
\end{equation}
yields
\begin{equation}
\mu\frac{\partial}{\partial t}{\mathbf{H}(\mathbf{r},t)}=-\nabla\times{\mathbf{E}}(\mathbf{r},t),\label{eq:induct}
\end{equation}
where $\nabla \times$ denotes the curl operator and $\mu$ the permeability of the medium under consideration. In the~case of RF for accelerators, often $\mu=\mu_0=\unit[4\,\pi\times10^{-7}]{Vs A^{-1}m^{-1}}$, since cavities are evacuated. Replacing the electric flux density with the electric field strength using the material relation
\begin{equation}
\varepsilon\,\mathbf{E}(\mathbf{r},t) =\mathbf{D}(\mathbf{r},t)
\end{equation}
and expressing the current densities as  sum of Ohmic current densities and current densities from excitation sources
\begin{equation}
\mathbf{J}(\mathbf{r},t) =\sigma\,\mathbf{E}(\mathbf{r},t) + \mathbf{J}_\mathrm{s}(\mathbf{r},t)
\end{equation}
in Amp\'ere's law delivers
\begin{equation}
\varepsilon\frac{\partial}{\partial t}{\mathbf{E}(\mathbf{r},t)}=-\sigma\,{\mathbf{E}(\mathbf{r},t)}+\nabla\times{\mathbf{H}(\mathbf{r},t)}-\mathbf{J}_\mathrm{s}(\mathbf{r},t).\label{eq:ampere}
\end{equation}
Here, $\varepsilon$ denotes the permittivity and $\sigma$ the conductivity.  In case of evacuated cavity structures, $\varepsilon=\varepsilon_0=\unit[8.854\times10^{-12}]{As V^{-1}m^{-1}}$ and $\sigma = \unit[0]{Sm^{-1}}$ are often chosen. The excitation current densities model excitations, such as currents assigned to waveguide ports or the current of charged particle beams. The~mutually coupled partial differential equations (\ref{eq:induct}) and (\ref{eq:ampere}) can be formally expressed using a "matrix" of differential operators as follows:
\begin{equation}
\begin{pmatrix}
\varepsilon & 0\\
0 & \mu
\end{pmatrix}
\frac{\partial}{\partial t}
\begin{pmatrix}
\mathbf{E}(\mathbf{r},t)\\
\mathbf{H}(\mathbf{r},t)
\end{pmatrix}
=
\begin{pmatrix}
-\sigma & \nabla\times\\
-\nabla\times & 0
\end{pmatrix}
\begin{pmatrix}
\mathbf{E}(\mathbf{r},t)\\
\mathbf{H}(\mathbf{r},t)
\end{pmatrix}+
\begin{pmatrix}
-1\\
0
\end{pmatrix}
\mathbf{J}_\mathrm{s}(\mathbf{r},t){.}\label{eq:contcurlcurl}
\end{equation}

\begin{figure}
  \centering
  \subfigure[]{\includegraphics[width=0.9\textwidth]{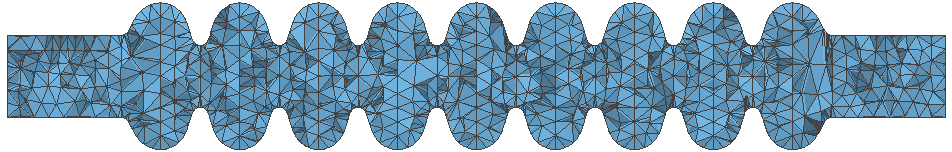}\label{fig:tet_mesh}}
  \subfigure[]{\includegraphics[width=0.90\textwidth]{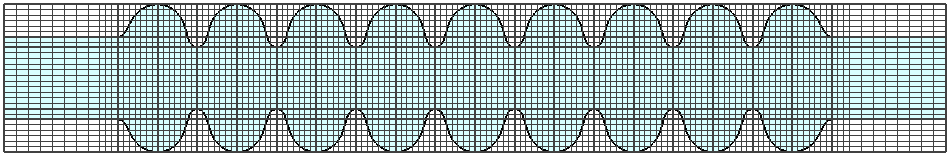}\label{fig:hex_mesh}}
  \caption{Cut plane view showing the volume discretization of a third-harmonic nine-cell TESLA cavity\cite{Vogel}. Subfigure (a) depicts the discretization based on a tetrahedral mesh, whereas subfigure (b) depicts the discretization based on a hexahedral mesh. Both plots are created with CST Studio Suite\textregistered~\cite{CST}.}
  \label{fig:meshing_example}
\end{figure}
Volume-based spatial discretization techniques, including FDTD, FIT, and FEM, play a particularly important role in modelling hollow waveguiding structures. They employ volumetric meshes (see Fig.~\ref{fig:meshing_example}) to discretize the domain, which allows (\ref{eq:contcurlcurl}) to be represented as a matrix-vector equation:
\begin{equation}
\mathbf{M}
\dwt
\mathbf{x}(t)
=
\mathbf{A}\,\mathbf{x}(t)+
\mathbf{B}\,\mathbf{i}(t)\textnormal{.}\label{eq:disccurlcurl}
\end{equation}
The choice of the mesh for discretization is crucial. Tetrahedral meshes, as shown in Fig.~\ref{fig:tet_mesh}, are very flexible to discretize sophisticated and complex geometries. In particular, tetrahedral elements with curved sides \cite[Section~4.9.1]{Jin} are employed to represent curved boundaries exactly or much more accurately. Hexahedral meshes, as illustrated in Fig.~\ref{fig:hex_mesh}, are inherently less flexible, posing challenges when modeling structures with boundaries that are not aligned with the mesh. In the worst case, this results in notorious (and poor) staircase approximations. Improved formulations that more accurately account for curved geometries using partially filled hexahedral mesh cells are proposed in Refs.~\cite{DeyMittra1997,ZagorodnovSchuhmannWeiland2003,SchreiberClemensVanRienen2004}. It is obvious that the creation of tetrahedral meshes is more demanding than the generation of hexahedral meshes. Generally, meshes should be selected to accurately represent the geometry. Furthermore, the spatial discretization must be sufficiently fine to resolve the electromagnetic wavelength. A commonly used guideline for FIT on hexahedral meshes requires the edge lengths to be smaller than $\unit[10]{\%}$ of the~shortest wavelength in the structure under study~\cite{Munteanu2010}. Note, however, that the exact requirement depends on the~numerical method and desired accuracy.

Equation (\ref{eq:disccurlcurl}) is semi-discrete because the explicit spatial derivatives have vanished, while the time derivatives are still present. Note that Eq.~(\ref{eq:contcurlcurl}) has a structure similar to Eq.~(\ref{eq:disccurlcurl}). Representations of Eq.~(\ref{eq:disccurlcurl}) resulting from the FIT can be found in e.g.\ Eqs.~(1) and (2) in \cite{Wittig1} or in Eq.~(5) in \cite{Wittig2}, while those arising from the FEM are provided e.g.\ in Ref.~\cite[Section~12.11.1]{Jin}.  Although FIT formulations on triangular meshes exist~\cite{vanRienen1999}, they typically use a hexahedral discretization. FEM formulations commonly employ tetrahedral discretizations. Note that FEM on hexahedral meshes is also available as shown in Ref.~\cite[Section~6.5.4]{rylander2012computational}. For historical reasons, the matrix $\mathbf{M}\in\mathbb{R}^{N\times N}$ is often referred to as mass matrix and the~matrix $\mathbf{A}$ as stiffness matrix. This wording results from FEM formulations, which have their origins in structural mechanics. The matrix $\mathbf{B}\in\mathbb{R}^{N\times N_\mathrm{t}}$ is sometimes designated as input matrix. The $N_\mathrm{t}$ columns of the input matrix contain a discrete representation of the excitation current densities referring to the $N_\mathrm{t}$ waveguide modes, which are defined at the waveguide ports. The modal excitation currents are listed in the  vector $\mathbf{i}(t)\in\mathbb{R}^{N_\mathrm{t}}$. The corresponding modal voltages are collated in the vector $\mathbf{v}(t)\in\mathbb{R}^{N_\mathrm{t}}$ and are typically determined by
\begin{equation}
\mathbf{v}(t) = 
\mathbf{C}\,
\mathbf{x}(t)
\label{eq:output}\textnormal{,}
\end{equation}
where $\mathbf{C}\in\mathbb{R}^{N_\mathrm{t}\times N}$ is the output matrix. The $N_\mathrm{t}$ rows of the output matrix contain the discrete electric field distributions of respective waveguide modes, defined at the waveguide ports. The reader is referred to Section~II~A.\ in Ref.~\cite{Flisgen_TMMT} for a discussion of waveguide port implementation, including the relationship between modal currents and voltages and the transverse magnetic and electric fields on the port cross section. For the treatment of waveguide ports in numerical methods, see Section~2 in Ref.~\cite{Wittig2} for the~FIT and Chapter~11.1 in Ref.~\cite{Jin} for the FEM.

The transient electric and magnetic fields in the domain are represented by the coefficients of the~time-dependent vector $\mathbf{x}(t)$. In this way, $\mathbf{x}(t)$ provides a discrete representation of the full electromagnetic field distribution. The $N$ degrees of freedom are often allocated at the midpoints of edges and facets of mesh cells. The number of degrees of freedom can become extremely large, ranging from $N\approx10^3$ up to $10^{10}$ or even higher. This is necessary to approximate the continuous form of Eq.~(\ref{eq:contcurlcurl}) by its spatially discretized counterpart~Eq.~(\ref{eq:disccurlcurl}). In other words, the large number of degrees of freedom arises from representing the continuous field distributions in a discrete form. This discrete form can be thought of as a quasi-continuum. Due to the local nature of spatial differential operators, each degree of freedom in the discretized domain interacts only with a small number of neighbouring degrees of freedom. Therefore, most matrix entries are zero, resulting in sparse matrices [see e.g.\ Fig.~\ref{fig:spy_plot}]. Sparse matrices are typically stored in specialized formats that record only the non-zero entries and their indices, thereby significantly reducing memory consumption.

In fact, Eqs.~(\ref{eq:disccurlcurl}) and (\ref{eq:output}) represent an impedance formulation, as modal voltages are related to modal currents. The equations provided in Section~II-E in Ref.~\cite{Flisgen_TMMT} allow for transferring this system into a~scattering formulation:
\begin{align}
\mathbf{M} \dwt \mathbf{x}(t) &=\underbrace{\left[\mathbf{A}-\mathbf{B}\mathbf{D}_z^{-1}\mathbf{C}\right]}_{\mathbf{\bar{A}}}\,\mathbf{x}(t)+ \underbrace{\sqrt{2}\mathbf{B}\mathbf{D}_z^{-1/2}}_{\mathbf{\bar{B}}}\,\mathbf{a}(t),\label{eq:states}\\
\mathbf{b}(t) &= \underbrace{\sqrt{2}\mathbf{D}_z^{-1/2}\mathbf{C}}_{\mathbf{\bar C}}\,\mathbf{x}(t)- \mathbf{I}\,\mathbf{a}(t)\textnormal{,}\label{eq:outputs}
\end{align}
where $\mathbf{a}(t)\in\mathbb{R}^{N_\mathrm{t}}$ and $\mathbf{b}(t)\in\mathbb{R}^{N_\mathrm{t}}$ are vectors collating the amplitudes of the incident and scattered waves, respectively. Moreover, $\mathbf{I}$ is the identity matrix and $\mathbf{D}_z$ a diagonal matrix holding the (in this case) constant impedances of the $N_\mathrm{t}$ waveguide modes.  The identity matrix multiplied by $-1$ acts as the~feedthrough or feedforward matrix. Both representations Eqs.~(\ref{eq:disccurlcurl})--(\ref{eq:output}) and Eqs.~(\ref{eq:states})--(\ref{eq:outputs}) are by no means unique--neither in FEM, FIT, nor other formulations--since state transformations can produce equivalent systems with different state vectors and matrices. 

\subsection{Time Domain Approaches\label{sec:time_do}}
This section describes the explicit treatment of time derivatives in the semi-discrete equations. Generally, time domain schemes have various advantages such as:
\begin{itemize}
 \item no limitation to sinusoidal excitations or steady-state operation, enabling the analysis of system responses to general transient signals (e.g., modulated or measured signals),
 \item non-linear electromagnetic problems can be modelled (see e.g.~Ref.~\cite{Klingbeil} for examples related to RF for accelerators),
 \item system responses over a wide frequency range are often readily available.
\end{itemize}

Formally multiplying the inverse of the mass matrix from the left to either Eq.~(\ref{eq:disccurlcurl}) or Eq.~(\ref{eq:states}) delivers a~classical linear initial value problem. For the scattering formulation (\ref{eq:states}) this reads as
\begin{equation}
\dwt
\mathbf{x}(t)
=
{\mathbf{M}}^{-1}\mathbf{\bar{A}}\,\mathbf{x}(t)+
{\mathbf{M}}^{-1}\mathbf{\bar{B}}\,\mathbf{a}(t)\textnormal{.}\label{eq:sss}
\end{equation}
In close analogy to control theory, this equation can be interpreted as a state equation with the state vector $\mathbf{x}(t)$. The state equation governs the evolution of the system state and, consequently, the field distribution within the considered volume over time. Indeed, Eq.~(\ref{eq:sss}) provides the change in the electric and magnetic field distributions at the initial time $t_0$, based on the state $\mathbf{x}(t_0)$ and the excitation $\mathbf{a}(t_0)$ at that moment. This allows for determining the system state $\mathbf{x}(t_1)$ at $t_1 = t_0 +\Delta t$, where $\Delta t$ is a~sufficiently small step size in time. Based on the now known state $\mathbf{x}(t_1)$ and the excitation $\mathbf{a}(t_1)$ at $t_1$, the state at $t_2 = t_1 +\Delta t$ is available. Following this iterative procedure, the system state $\mathbf{x}(t_k)$ at $t_k = t_0 +k\,\Delta t$ based on the~initial state $\mathbf{x}(t_0)$ and the transient excitation $\mathbf{a}(t)$ can be computed. Formally, the relationship between the~states at two consecutive time steps $t_{k+1}$ and $t_{k}$ is obtained by integrating Eq.~(\ref{eq:sss}) with respect to time:
\begin{equation}
\mathbf{x}(t_{k+1}) = \mathbf{x}(t_k+\Delta t)=\int_{t_k}^{t_k+\Delta t}\,{\mathbf{M}}^{-1}\mathbf{\bar{A}}\,\mathbf{x}(\tau)+
{\mathbf{M}}^{-1}\mathbf{\bar{B}}\,\mathbf{a}(\tau)\,\mathrm{d}\tau+\mathbf{x}(t_k)\textnormal{.}\label{eq:update}
\end{equation}
Numerous numerical methods have been proposed to evaluate the next state $\mathbf{x}(t_{k+1}) $ based on the current state and previous states; see, for example, Refs.~\cite{GriffithsHigham2010,HairerNorsettWanner1993} and the references therein. Single-step methods, such as Forward Euler, Backward Euler, Crank–Nicolson or Runge–Kutta, compute the next state using only the current state. In contrast, multistep methods, such as the Adams–Bashforth method, the~Adams–Moulton method or the Leapfrog method, use the current state together with one or more preceding states. Explicit methods, such as Forward Euler, Runge–Kutta, Adams–Bashforth or Leapfrog compute the next state directly from known quantities at the current or previous steps and the costly solution of a system of linear equations at every time step is not required. However, the largest time step $\Delta t$ has to be below a certain threshold for a stable iteration. The shortest edge in the mesh often dictates the longest stable time step for explicit schemes. This is particularly challenging when studying structures that contain small but non-negligible geometrical features, such as higher-order mode couplers with a dedicated notch-filter effect \cite{Papke2017}. The implicit methods Backward Euler, Crank–Nicolson, and Adams–Moulton compute the next state using information that depends on this unknown next state, which requires solving a system of linear equations at each time step. While this increases computational effort, implicit methods largely circumvent the severe stability restrictions of explicit schemes. After computing the field distributions $\mathbf{x}(t_k)$ with one of the schemes described above, the transient scattered wave amplitudes at time $t_k$ are directly determined using the output equation
\begin{equation}
\mathbf{b}(t_k)
=
\mathbf{\bar{C}}\,\mathbf{x}(t_k)+
\mathbf{\bar{D}}\,\mathbf{a}(t_k).
\end{equation}
This discussed procedure yields both the transient field distributions $\mathbf{x}(t_k)$ and the transient scattered wave amplitudes $\mathbf{b}(t_k)$ at discrete time steps $t_k$ based on a given sampled transient excitation $\mathbf{a}(t_k)$.

It should be noted that explicit time-domain iterations are rarely practical in standard FEM formulations. Equations (\ref{eq:sss}) and (\ref{eq:update}) require the inverse of the typically large, sparse, non-diagonal mass matrix $\mathbf{M}$. This inverse is not readily available. Computing $\mathbf{M}^{-1}$ is expensive as the number of required floating point operations scales with $N^3$. Furthermore, matrix inversion can be numerically unstable and may lead to fill-in. Fill-in refers to the phenomenon in which a matrix that is initially sparse (i.e., contains mostly zero entries) develops additional nonzero elements during the execution of an algorithm, thereby reducing its sparsity. Instead of inverting the mass matrix, one could solve a system of linear equations at each time step; however, this undermines the main advantage of explicit methods, which is to avoid such solves. Despite these facts, modifications of FEM formulations allowing for explicit time-domain iterations exist. For instance, mass lumping (see e.g. Refs.~\cite{bathe1996finite,rylander2012computational,egger2021second}) approximates the mass matrix with a diagonal matrix, allowing its inverse to be obtained simply by inverting the diagonal elements. Further formulations such as the so-called Discontinuous Galerkin Finite-Element Method (DG-FEM) (see e.g. Refs.~\ \cite{schnepp2012efficient, hesthaven2008nodal} and references therein) emerged, allowing for efficient explicit time-marching schemes based on FEM.

In contrast to standard FEM formulations, the FIT on hexahedral meshes naturally yields a diagonal mass matrix $\mathbf{M}$. Consequently, its inversion is trivial, making explicit time-stepping schemes efficient and practical. Combining the explicit leapfrog scheme with the FIT on hexahedral meshes leads to the~well-known FDTD method. The FDTD algorithm was first introduced by K. S. Yee in 1966~\cite{yee1966numerical} and numerous extensions and modifications have been proposed since then. A comprehensive description of this method is provided in Ref.~\cite{taflove2000computational} and in references therein. Despite its age, FDTD remains an important and widely used approach for electromagnetic simulations. It is well suited for problems whose characteristic lengths are comparable to or exceed the lengths of the electromagnetic waves. This is typically the case for simulations arising from microwave or optical applications, such as, e.g.\ Ref.~\cite{phung2019microwave} or Ref.~\cite{rahimof2024study}, respectively.

\begin{figure}
 \centering 
  \subfigure[]{\includegraphics[width=0.48\textwidth]{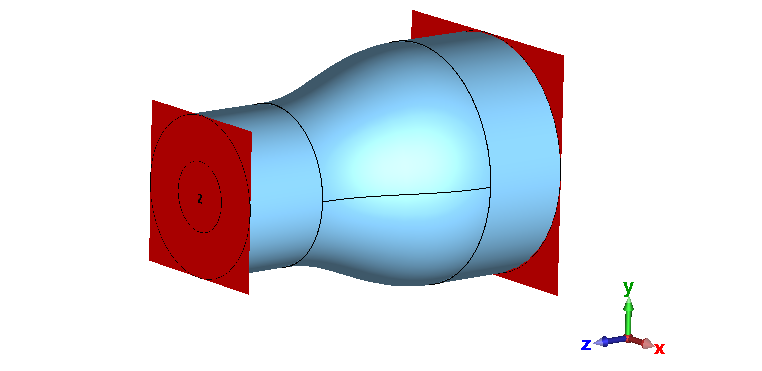}\label{fig:struct3D}}
  \subfigure[]{\includegraphics[width=0.48\textwidth]{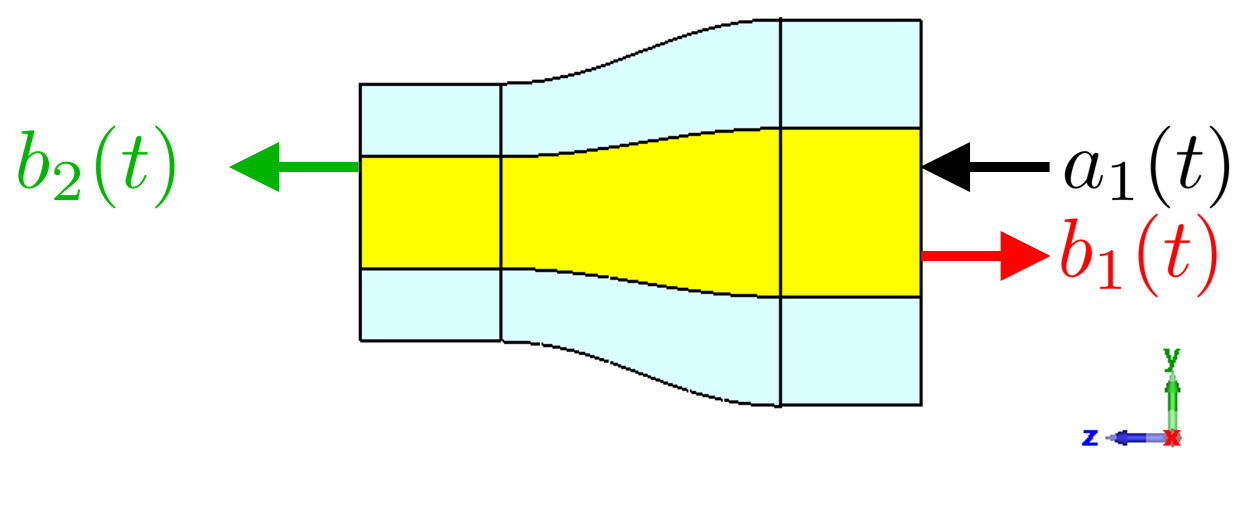}\label{fig:struct2D}}\\
  \subfigure[]{\includegraphics[width=0.48\textwidth]{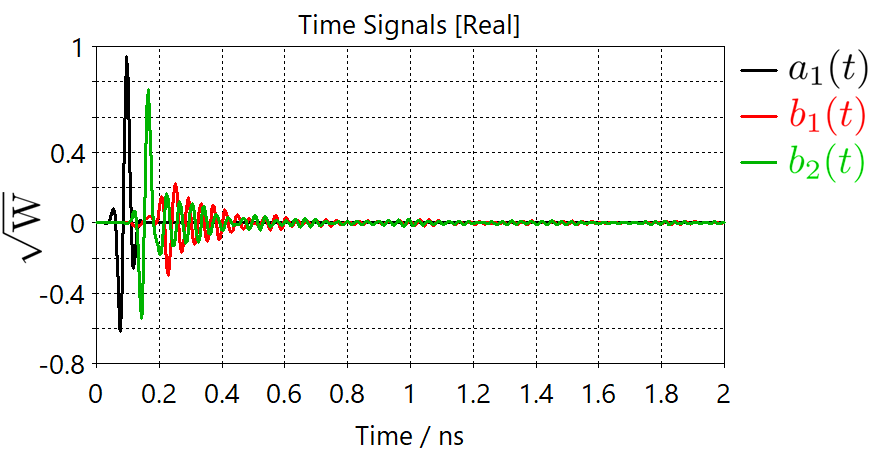}\label{fig:transient_signals}}
  \subfigure[]{\includegraphics[width=0.48\textwidth]{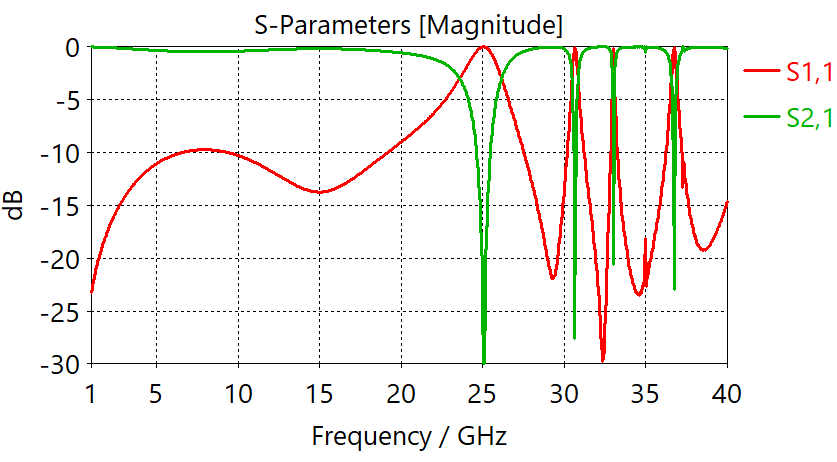}\label{fig:xxz}}
  \caption{Time-domain methods for determination of frequency-domain quantities using a coaxial waveguide taper as example. (a) Perspective view of the taper. Thered areas represent the waveguide ports, where exclusively the~transverse electromagnetic mode is considered. (b) Slice view of the taper. The inner conductor is shown in yellow, while the dielectric is shown in light blue. Port $1$ (left-hand side) is excited with the incident signal $a_1(t)$. The~backscattered signal into Port $1$ is denoted by $b_1(t)$, whereas the signal transmitted into Port $2$ is denoted by $b_2(t)$. (c)~Transient excitation $a_1(t)$ as black curve and transient system responses in red and green, respectively. (d)~Magnitude of taper reflection and transmission coefficients, derived from the transient signals shown in~(c).}
  \label{fig:sparam}
\end{figure}
Time-domain methods are very often employed to compute broadband frequency-domain properties, such as scattering parameters, within a single simulation run~\cite{Munteanu2010}. Figure~\ref{fig:sparam} illustrates the required procedure through an example. Figure~\ref{fig:struct3D} shows a coaxial waveguide taper in a perspective view. The~red facets mark the waveguide ports used to excite the transverse electromagnetic (TEM) mode. The right port is referred to as Port~$1$, while the left port is referred to as Port~$2$. Figure~\ref{fig:struct2D} provides a slice view of the structure, with incident and scattered wave amplitudes assigned to the two ports. In a first step, Port~$1$ is excited in time domain with a Gaussian pulse. This pulse is depicted in Fig.~\ref{fig:transient_signals} in black. The~spectrum of the Gaussian pulse is such that it carries root-mean-square energy in the frequency range from $f_\mathrm{min}$ to $f_\mathrm{max}$. These two frequencies define the lower and upper bounds for the computation of the scattering parameters. The described time-domain iterations are now used to determine the discrete signal $b_1(t_k)$ reflected into Port $1$ as well as the discrete signal $b_2(t_k)$ transmitted/scattered into Port~$2$. Figure~\ref{fig:transient_signals} presents the responses to the stimulus, shown as solid red and solid green curves, respectively. The scattering coefficients resulting from the Port 1 excitation are readily determined in a post-processing step by dividing the fast Fourier transform of the system responses by that of the excitation:
\begin{equation}
\underline{s}_{11}(\omega_k)=\frac{\mathrm{FFT}[b_1(t_k)]}{\mathrm{FFT}[a_1(t_k)]}, \quad\underline{s}_{21}(\omega_k)=\frac{\mathrm{FFT}[b_2(t_k)]}{\mathrm{FFT}[a_1(t_k)]},
\end{equation}
where $\mathrm{FFT}[\quad]$ is the fast Fourier transform of a discrete signal. The obtained complex-valued reflection and transmission coefficients $\underline{s}_{11}(\omega_k)$ and $\underline{s}_{21}(\omega_k)$ are sampled on the discrete angular frequencies $\omega_k$. The remaining scattering coefficients $\underline{s}_{12}(\omega_k)$ and $\underline{s}_{22}(\omega_k)$ of the example structure are obtained in a~similar way by computing the transient system responses resulting from an excitation at Port~$2$ with the~Gaussian pulse.

Ideally, the time-marching schemes should be executed until all signals have fully decayed to zero~\cite{Munteanu2010}. Practically, however, this is impossible, as it would require an infinite number of discrete time steps. Therefore, time-domain iterations are typically terminated once the energy in the computational domain has dropped below a user-defined threshold or a maximum number of time steps has been reached. The residual energy in the computational domain after termination of the time-domain iteration is an important quantity to monitor: if it does not fall below a specified threshold, ripples will appear in the scattering curves. These ripples are numerical artifacts and must not be interpreted as inherent properties of the structure under investigation.

\begin{figure}
  \centering
  \subfigure[]{\includegraphics[width=0.45\textwidth]{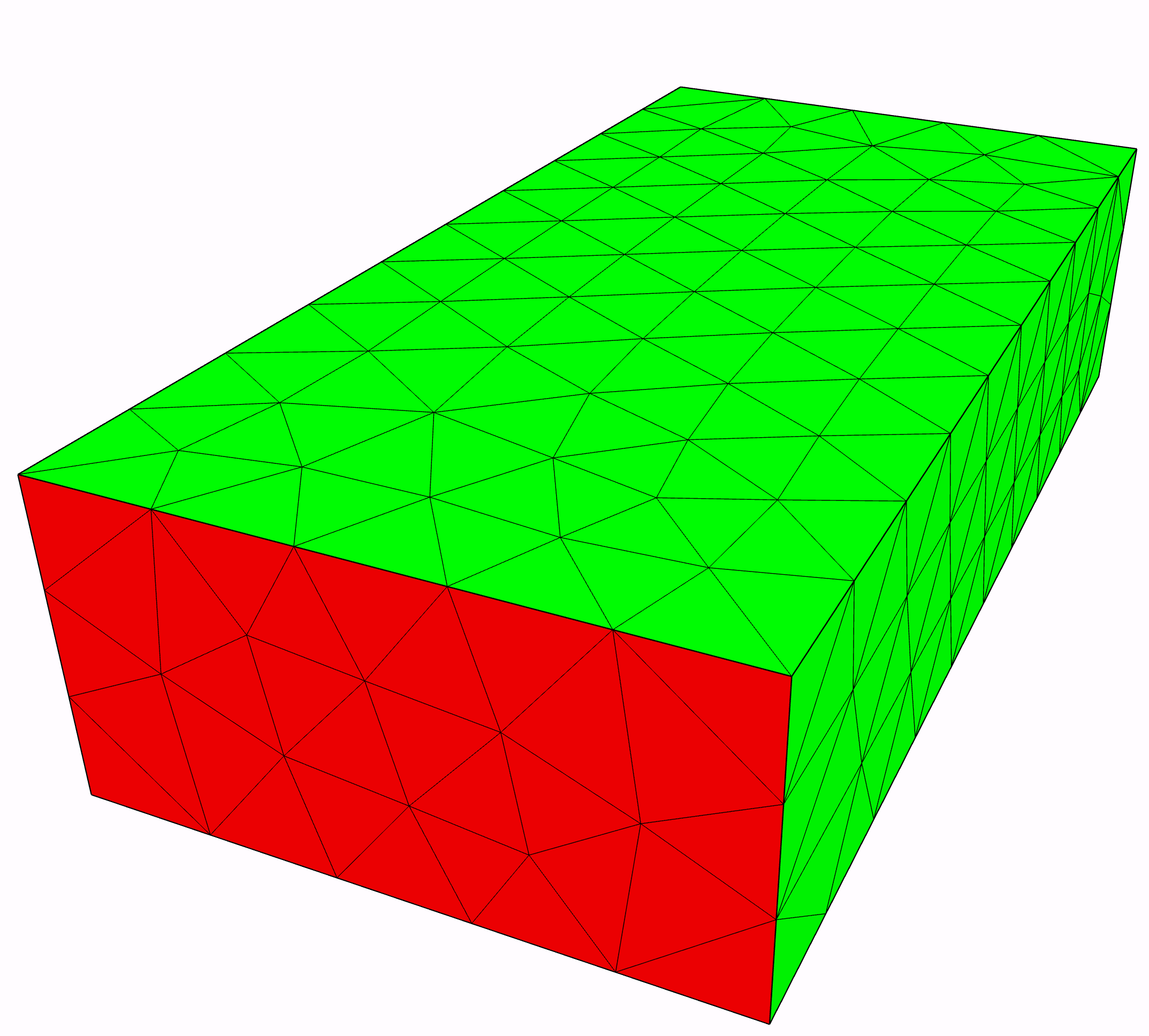}\label{fig:traingulated_wg}}\quad
  \subfigure[]{\includegraphics[height=0.4\textwidth]{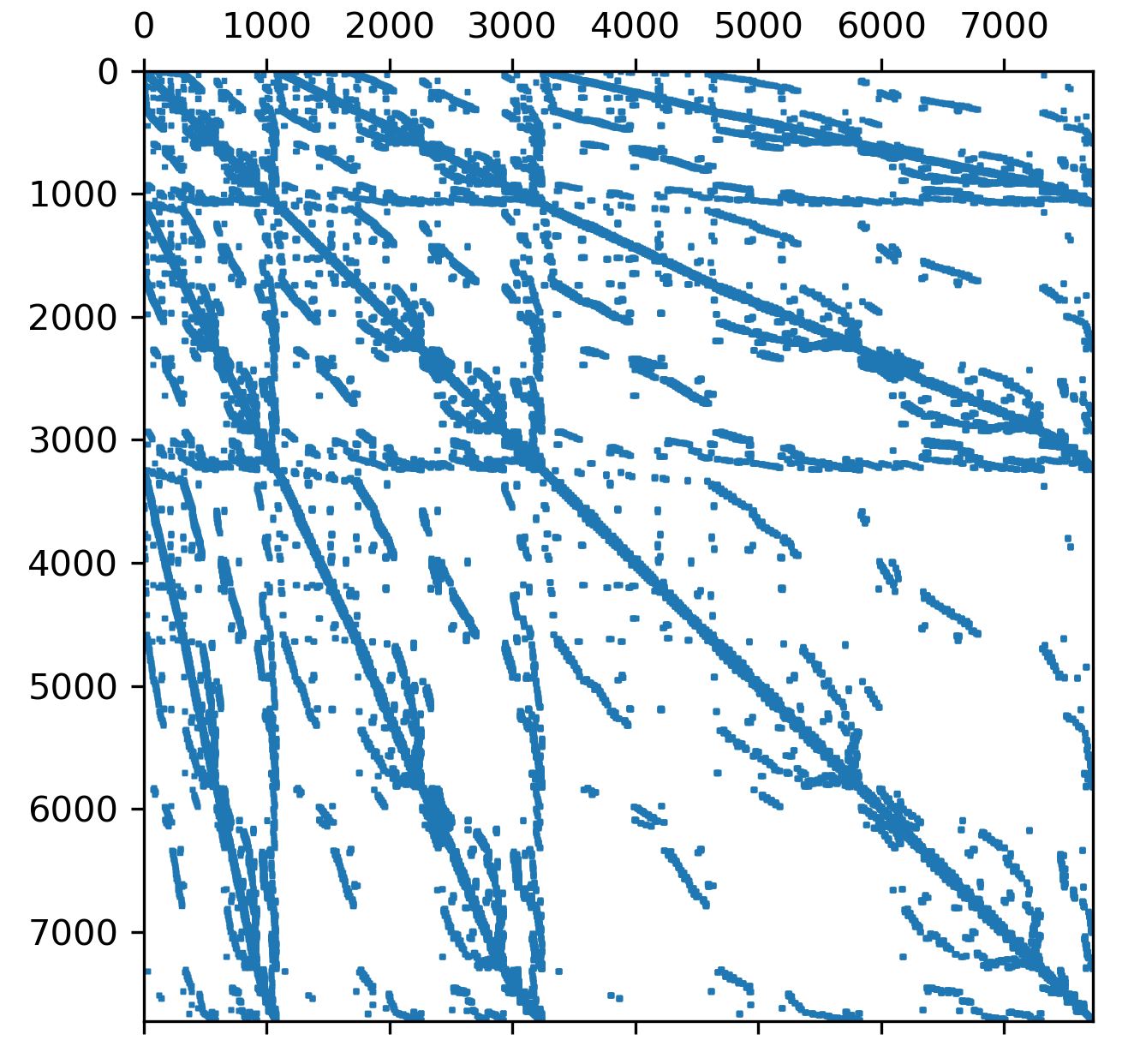}\label{fig:spy_plot}}
  \caption{(a) Rectangular waveguide discretized using a tetrahedral mesh as an example. A waveguide port for excitation is depicted as red facet. Another port is located at the rear of the structure and is not visible in the plot. (b) Spy plot of the matrix $\underline{\mathbf{G}}$ visualizing its sparsity pattern. The matrix $\underline{\mathbf{G}}$ is obtained from the discretization of the waveguide in (a) using FEM with second-order basis functions.}
  \label{fig:freqdo_example}
\end{figure}
\subsection{Frequency Domain Approaches\label{sec:freq_do}}
Frequency domain methods allow for a steady-state analysis. They are capable of determining system responses due to sinusoidal excitations. As is customary in the frequency domain, these time-dependent quantities are represented as phasors. For example, this yields the following expressions for the semi-discrete impedance formulation in Eqs.~(\ref{eq:disccurlcurl}) and (\ref{eq:output}):
\begin{align}
\mathbf{i}(t) &= \Re\left[{\underline{\mathbf{i}}(j\omega)}\right],\quad
\underline{\mathbf{i}}(j\omega) = \underline{\mathbf{i}} \exp{(j\omega t)},\\
\mathbf{x}(t) &= \Re\left[{\underline{\mathbf{x}}(j\omega)}\right],\quad
\underline{\mathbf{x}}(j\omega) = \underline{\mathbf{x}} \exp{(j\omega t)},\\
\mathbf{v}(t) &= \Re\left[{\underline{\mathbf{v}}(j\omega)}\right],\quad
\underline{\mathbf{v}}(j\omega) = \underline{\mathbf{v}} \exp{(j\omega t)},
\end{align}
where $\Re\left[\quad\right]$ denotes the real part of a complex number, $ \exp{(\quad)}$ the natural exponential function, $j$ the~imaginary unit, and $\omega$ the angular frequency. The phasors of the modal excitations currents, the~system state (i.e.\ the field distribution), and the modal voltages are given by $\underline{\mathbf{i}}\in\mathbb{C}^{N_\mathrm{t}}$, $\underline{\mathbf{x}}\in\mathbb{C}^{N}$, $\underline{\mathbf{v}}\in\mathbb{C}^{N_\mathrm{t}}$, respectively. The phasors carry amplitude and phase information. The restriction to harmonic (sinusoidal) excitations at a single angular frequency allows for transforming the differential Eq.~(\ref{eq:disccurlcurl}) into an algebraic equation involving complex-valued quantities:
\begin{equation}
j\omega\mathbf{M}
\underline{\mathbf{x}}
=
\mathbf{A}\,\underline{\mathbf{x}}+
\mathbf{B}\,\underline{\mathbf{i}}\textnormal{.}\label{eq:Zstatefd}
\end{equation}
This statement can be rearranged to
\begin{equation}
\underbrace{\left[j\omega\mathbf{M}
-\mathbf{A}\right]}_{\underline{\mathbf{G}}}\,\underline{\mathbf{x}}
=
\underbrace{\mathbf{B}\,\underline{\mathbf{i}}}_{\underline{\mathbf{b}}}\textnormal{,}\label{eq:lse}
\end{equation}
which forms a system of linear equations with a frequency-dependent matrix $\underline{\mathbf{G}}\in\mathbb{C}^{N\times N}$ and a constant right-hand side $\underline{\mathbf{b}}\in\mathbb{C}^N$. The matrix $\underline{\mathbf{G}}$ is large, and sparse, as $\mathbf{M}$ and $\mathbf{A}$ are themselves large sparse matrices.  Figure~\ref{fig:traingulated_wg} shows the volume discretization of a rectangular waveguide using a tetrahedral mesh as an example. Both ends of the structure are equipped with waveguide ports accounting solely for the TE$_{10}$-mode. Figure~\ref{fig:spy_plot} presents the spy plot of the matrix $\underline{\mathbf{G}}$. This matrix is obtained from the discretization of the rectangular waveguide. A spy plot is a graphical representation of a matrix that displays the locations of its nonzero entries, thereby visualizing the matrix’s sparsity pattern. In the~figure, blue dots indicate the nonzero entries.
 
The unknown field distributions in Eq.~(\ref{eq:lse}) arise from the excitation $\underline{\mathbf{i}}$ and are formally given by
\begin{equation}
\underline{\mathbf{x}}
=
\underbrace{\left[j\omega\mathbf{M}
-\mathbf{A}\right]^{-1}}_{\underline{\mathbf{G}}^{-1}}\underbrace{\mathbf{B}\,\underline{\mathbf{i}}}_{\underline{\mathbf{b}}}\textnormal{.}\label{eq:inverse}
\end{equation}
As mentioned in Section~\ref{sec:time_do}, explicitly computing the inverse of a large sparse matrix is often impractical: the required number of floating-point operations scales with $N^3$, matrix inversion can be numerically unstable, and may introduce fill-in that destroys sparsity. Instead, iterative methods, such as Conjugate Gradient (CG), Minimal Residual (MINRES), Generalized Minimal Residual (GMRES) or BiCG (Biconjugate Gradient), are often employed to find the unknown state vector in Eq.~$(\ref{eq:lse})$ or Eq.~($\ref{eq:inverse}$). Details on these methods can be found in Ref.~\cite{meister2026numerical} and in references therein. Iterative methods require an initial guess $\underline{\mathbf{x}}^{(0)}$ of the exact solution $\underline{\mathbf{x}}^*$ (the exact solution satisfies $\underline{\mathbf{G}}\,\underline{\mathbf{x}}^{*}=\underline{\mathbf{b}}$). In the absence of a better initial guess, the iteration may be initialized with the zero vector: $\underline{\mathbf{x}}^{(0)}=\mathbf{0}$. Based on the initial guess $\underline{\mathbf{x}}^{(0)}$, the~matrix $\underline{\mathbf{G}}$, and the right-hand side $\underline{\mathbf{b}}$, an improved approximation $\underline{\mathbf{x}}^{(1)}$ of the exact solution is determined in a first iteration step:
\begin{equation}
\underbrace{\|\underline{\mathbf{x}}^{(0)}-\underline{\mathbf{x}}^{*}\|}_{e^{(0)}}>\underbrace{\|\underline{\mathbf{x}}^{(1)}-\underline{\mathbf{x}}^{*}\|}_{e^{(1)}}\textnormal{.}
\end{equation}
Here, $\|\quad\|$ is a vector norm and $e^{(k)}$ the error in the $k$th iteration step. The new improved vector is obtained from the previous vector by an update formula of the form:
\begin{equation}
\underline{\mathbf{x}}^{(k+1)} = \mathrm{func}\left[\underline{\mathbf{G}},\underline{\mathbf{b}},\underline{\mathbf{x}}^{(k)}\right]\textnormal{,}
\end{equation}
where $\mathrm{func}[\quad]$ depends on the specific iteration method chosen. As the number of iterations approaches infinity, the scheme is designed to converge to the exact solution, such that the error tends to zero:
\begin{equation}
\lim_{k \to \infty} e^{(k)} = \lim_{k \to \infty}\|\underline{\mathbf{x}}^{(k)}-\underline{\mathbf{x}}^{*}\|=0\textnormal{.}
\end{equation}
Note, however, that convergence is not guaranteed and depends on the choice of an appropriate iterative method, which should be selected according to the properties of the matrix $\underline{\mathbf{G}}$.

In practical applications, the number of iterations must be finite. Therefore, the iteration is stopped once the approximation achieves a sufficient level of accuracy. Stopping criteria based on the error $e^{(k)}$ are not practical, as they require knowledge of the exact solution $\underline{\mathbf{x}}^{*}$. However, this solution is not available. If it were known, the use of an iterative method would be unnecessary. Instead, stopping criteria based on the residual are employed. The residuum in the $k$th iteration step is given by
\begin{equation}
\underline{\mathbf{r}}^{(k)} =\underline{\mathbf{b}}-\underline{\mathbf{G}}\,\underline{\mathbf{x}}^{(k)}\textnormal{.}
\end{equation}
It follows directly from the definition that the exact solution $\underline{\mathbf{x}}^{*}$ is not required and that the residual tends to the zero vector as the approximation converges to the exact solution. Often relative residuals $\|\underline{\mathbf{r}}^{(k)}\|/\|\underline{\mathbf{r}}^{(0)}\|$ are chosen as stopping criterion for the iteration. Note, however, that small relative residuals do not necessarily result in a small relative error of the iterative solution. Theoretical considerations deliver
\begin{equation}
\frac{\|\underline{\mathbf{x}}^{*}-\underline{\mathbf{x}}^{(k)}\|}{\|\underline{\mathbf{x}}^{*}\|}\leq
\underbrace{\|\underline{\mathbf{G}}\|\cdot\|\underline{\mathbf{G}}^{-1}\|}_{\mathrm{cond}(\underline{\mathbf{G}})}\cdot\frac{\|\underline{\mathbf{r}}^{(k)}\|}{\|\underline{\mathbf{b}}\|}\textnormal{,}\label{eq:inequality}
\end{equation}
where $\|\underline{\mathbf{G}}\|$ is the norm of the matrix $\underline{\mathbf{G}}$ and $\|\underline{\mathbf{r}}^{(k)}\|/\|\underline{\mathbf{b}}\|$ represents the relative residual corresponding to the choice of the zero vector as the initial guess, since $\underline{\mathbf{r}}^{(0)} = \underline{\mathbf{b}}$ for $\underline{\mathbf{x}}^{(0)}=\mathbf{0}$. Clearly, Eq.~(\ref{eq:inequality}) defines an~upper limit for the relative error of the approximation $\underline{\mathbf{x}}^{(k)}$. This limit, and thus the relative error, may still be large even if the relative residual is small, particularly when the so-called condition number $\mathrm{cond}(\underline{\mathbf{G}})$ of the matrix $\underline{\mathbf{G}}$ is large. The condition number is not available in practical applications, because it requires the computation of a matrix inverse. The matrix is well-conditioned if $\mathrm{cond}(\underline{\mathbf{G}})\approx 1$ and ill-conditioned if $\mathrm{cond}(\underline{\mathbf{G}})>>1$. Ill-conditioned system matrices arise when large aspect ratios are present, for example when the length of the mesh edges or the material properties differ greatly. In practical computations, the relative residual will not decrease further below a certain level after a~sufficient number of iterations. Instead, it will stagnate at a level determined by the accuracy of the floating-point arithmetic while iterating. Although this level may appear small, e.g.\ $\|\underline{\mathbf{r}}^{(k)}\|/\|\underline{\mathbf{r}}^{(0)}\|\approx 10^{-12}$, the relative error of the iterative solution may still be large in case of an ill-conditioned matrix with $\mathrm{cond}(\underline{\mathbf{G}})\geq 10^{11}$. Therefore, it is very challenging to solve systems of linear equations with ill-conditioned system matrices.

Preconditioning is often used in this context as it allows for a reduction of the condition number of the involved matrix. Left-hand side preconditioning is conducted by multiplying Eq.~(\ref{eq:lse}) from the~left-hand side with a matrix $\underline{\mathbf{L}}_\mathrm{c}$:
\begin{equation}
\underline{\mathbf{L}}_\mathrm{c}\,\underline{\mathbf{G}}\,\underline{\mathbf{x}}
=\underline{\mathbf{L}}_\mathrm{c}\,\underline{\mathbf{b}}\textnormal{.}\label{eq:lhs_pre}
\end{equation}
The matrix $\underline{\mathbf{L}}_\mathrm{c}$ is chosen such that it is a good approximation of the (not available) inverse of $\underline{\mathbf{G}}$. Right-hand side preconditioning is performed by defining a further matrix $\underline{\mathbf{R}}_\mathrm{c}$ and a new state $\underline{\mathbf{y}}$. The relationship between new and old state is given by
\begin{equation}
\underline{\mathbf{x}} =\underline{\mathbf{R}}_\mathrm{c}\, \underline{\mathbf{y}}\textnormal{.}\label{eq:rhs_pre}
\end{equation}
Combining Eqs.~(\ref{eq:lhs_pre}) and (\ref{eq:rhs_pre}) yields the preconditioned system of linear equations
\begin{equation}
\underline{\mathbf{L}}_\mathrm{c}\,\underline{\mathbf{G}}\,\underline{\mathbf{R}}_\mathrm{c}\, \underline{\mathbf{y}}
=\underline{\mathbf{L}}_\mathrm{c}\,\underline{\mathbf{b}}\textnormal{.}\label{eq:cond}
\end{equation}
The choice of the so-called preconditioners $\underline{\mathbf{L}}_\mathrm{c}$ and $\underline{\mathbf{R}}_\mathrm{c}$ aims at $\mathrm{cond}(\underline{\mathbf{L}}_\mathrm{c}\,\underline{\mathbf{G}}\,\underline{\mathbf{R}}_\mathrm{c})<<\mathrm{cond}(\underline{\mathbf{G}})$, so that the better-conditioned problem (\ref{eq:cond}) is solved instead of the ill-conditioned problem (\ref{eq:lse}). Once $\underline{\mathbf{y}}$ is determined by (iteratively) solving (\ref{eq:cond}), the sought vector $\underline{\mathbf{x}}$ is readily determined by the mapping (\ref{eq:rhs_pre}). Section~5 in \cite{meister2026numerical} provides further information on preconditioning. 

After the determination of the field distributions, the impedance parameters are readily available from a frequency-domain representation of the output Eq.~(\ref{eq:output}). Formally combining Eq.~(\ref{eq:output}) with Eq.~(\ref{eq:inverse}) yields the transfer function
\begin{equation}
\underline{\mathbf{v}} = \underbrace{\mathbf{C}\,\left[j\omega\mathbf{M}
-\mathbf{A}\right]^{-1}\mathbf{B}}_{\underline{\mathbf{Z}}(j\omega)}\,\underline{\mathbf{i}}\textnormal{.}\label{eq:Zmat}
\end{equation}
Clearly, $\underline{\mathbf{Z}}(j\omega)\in\mathbb{C}^{N_\mathrm{t}\times N_\mathrm{t}}$ is the frequency-dependent impedance matrix of the structure under study. This matrix can be converted into a scattering matrix using well-established transfer relations in a post-processing step (see, e.g., Chapter~4 in~\cite{PozarMicrowave}). Alternatively, the scattering formulation Eqs.~(\ref{eq:states}) and (\ref{eq:outputs}) can be transferred directly to frequency domain to obtain a transfer function relating scattered and incident wave amplitudes:
\begin{equation}
\underline{\mathbf{b}} = \underbrace{\left[\mathbf{\bar{C}}\,\left[j\omega\mathbf{M}
-\mathbf{\bar{A}}\right]^{-1}\mathbf{\bar{B}}-\mathbf{I}\right]}_{\underline{\mathbf{S}}(j\omega)}\,\underline{\mathbf{a}}\textnormal{.}\label{eq:Smat}
\end{equation}

\subsection{Simulation of Resonant Structures\label{sec:highQ}}
\begin{figure}
  \centering
  \subfigure[]{\includegraphics[width=0.4\textwidth]{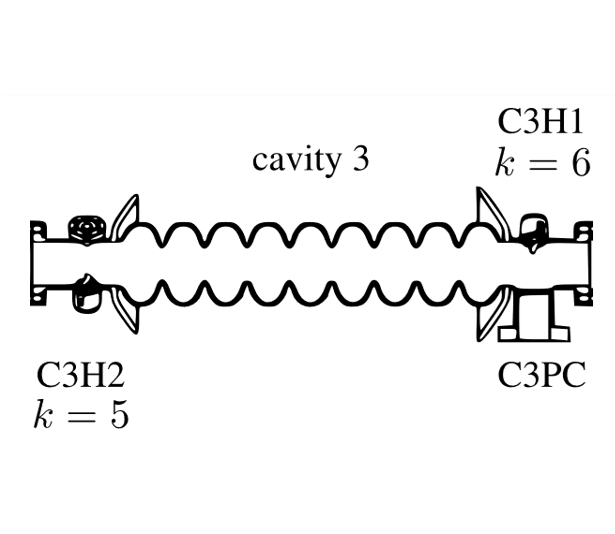}\label{fig:cavity}}\quad\quad
  \subfigure[]{\includegraphics[width=0.4\textwidth]{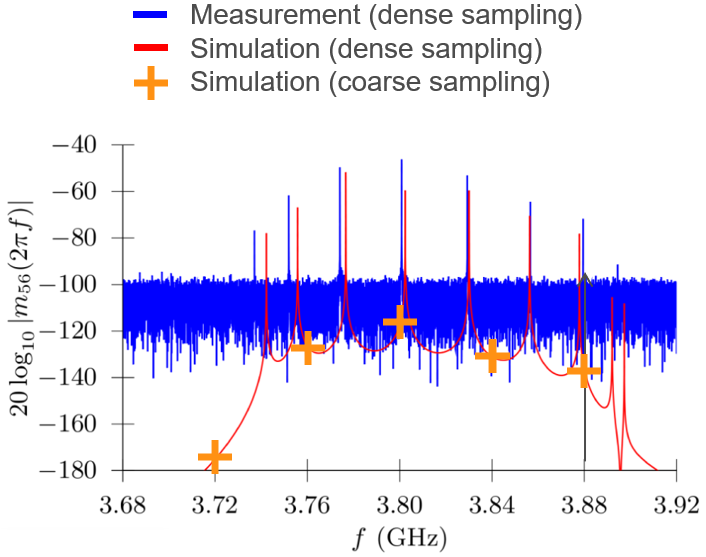}\label{fig:scavity_spec}}
  \caption{(a) Third-harmonic TELSA cavity with higher-order mode couplers on its left-hand side and higher-order mode and input coupler on the right-hand side. The illustration is copied from Ref.~\cite{Vogel}. (b) Absolute value of the~transmission from left to right higher-order mode coupler of the cavity depicted in (a). The measured transmission with a dense frequency-domain sampling is depicted in blue, simulated transmission with dense sampling in red and with coarse sampling in orange. The diagram is a modification of the Fig.~12(e) in Ref.~\cite{FlisgenACC39}.}
  \label{fig:cavity_and_spec}
\end{figure}

The simulation of scattering parameters of highly resonant structures is a challenging task. Highly resonant structures, such as elliptical shaped cavities shown in Figs.~\ref{fig:figuresofmerit},~\ref{fig:meshing_example}, and~\ref{fig:cavity}, are very important geometries in the context of RF for accelerators. Highly resonant structures support resonances with large quality factors
\begin{equation}
Q_m=\frac{\tilde{\omega}_m W_{m}}{P_{m}}\textnormal{.}
\end{equation}
Here, $\tilde{\omega}_m$ is the angular resonant frequency, $W_{m}$ the energy stored in the fields, and $P_{m}$ the power dissipated. The subscript $m$ denotes that all quantities refer to the $m$th resonance in the resonator. Often, various kinds of quality factors are considered. Intrinsic quality factors $Q_{0,m}$ exclusively account for intrinsic losses $P_{0,m}$ in the cavity. Intrinsic losses are typically attributed to surface losses resulting from the finite conductivity of the cavity boundaries. External quality factors $Q_{\mathrm{ext},m}$ exclusively account for external losses $P_{\mathrm{ext},m}$ in the cavity. External losses are related to the flow of energy out of the resonator via waveguide ports. Total quality factors $Q_{\mathrm{tot},m}$ account for intrinsic losses $P_{0,m}$ and external losses $P_{\mathrm{ext},m}$ in the cavity. Modes with a weak coupling to the waveguide ports are referred to as trapped modes. Trapped modes are characterized by a small $P_{\mathrm{ext},m}$ and thus by large $Q_{\mathrm{ext},m}$. The reader is referred to~Ref.~\cite{SchuhmannWeiland2000} for a rigorous analysis of trapped modes in accelerating cavities.

Resonant modes with large quality factors do have long rise times and decay times of field energy, i.e.\ their energy decay is proportional to $\exp(-\tilde{\omega}_m t/Q_m)$. Modes in resonators with copper walls may have $Q_m \approx 10^4$, whereas $Q_m \approx 10^{10}$ or even higher can be reached in case of superconducting walls~\cite{Padamsee2017}. The stored energy decayed to about $\unit[37]{\%}$ of its initial value after $Q_m / (2\pi)$ periods. Following~Eq.~(3) in~Ref.~\cite{SchuhmannWeiland2000}, a typical discretization in time domain with $20$ time steps per period requires $10^5$ to $10^{11}$ discrete time steps for normal and superconducting cavities, respectively. Owing to this large number of steps, time-domain approaches are not well-suited for the analysis of structures with high quality factors.

Frequency approaches inherently assume a steady state. Therefore, they do not suffer from long transients resulting from large quality factors. However, high quality factors result in sharp resonance peaks in spectra: $\Delta f_m =\tilde{f}_m/Q_m$, where $\Delta f_m$ is the width of the peak and $\tilde{f}_m$ the resonant frequency of the $m$th mode, respectively. Figure~\ref{fig:cavity} shows a sketch of a third-harmonic TESLA cavity (TM$_{01}$-$\pi$-mode at $\unit[3.9]{GHz}$) with higher-order mode and input couplers as an example. Figure~\ref{fig:scavity_spec} shows the~measured and simulated magnitude of the transmission between the two higher-order mode couplers. The~quality factors of this fundamental passband are governed by external losses and are in the~order of $10^6$ to $10^7$ as illustrated by Fig.~12 in Ref.~\cite{FlisgenACC39}.  The transmission spectrum needs to be computed on a sufficient number of frequency samples to properly resolve the resonant peaks. A too coarse frequency sampling [see orange crosses in Fig.~\ref{fig:scavity_spec}] does not provide a reasonable resolution. The distance between two frequency-samples should not be larger than $\Delta f_n/10\approx \unit[400]{Hz}$. The step size leads to approximately $6\times 10^5$ frequency samples in the interval from $\unit[3.68]{GHz}$ to $\unit[3.92]{GHz}$. Each sample of the transmission spectrum requires the computationally expensive solution of a large, sparse, and generally complex-valued system of linear equations, as discussed in Section~\ref{sec:freq_do}.  Consequently, straightforward frequency domain approaches are also not well-suited to analyze structures with large quality factors.

Instead, model-order reduction (MOR) techniques are the recommended approach to efficiently determine the network matrices of highly resonant structures. An overview on MOR from a system-theoretic perspective is provided in Ref.~\cite{Baur2014MOR}, whereas Ref.~\cite{Simeoni2009MOR} gives an overview on MOR with the focus on linear electromagnetic problems. MOR for FIT formulations is covered in~Refs.~\cite{Wittig1,Wittig2,SchuhmannWeiland2004}, while MOR for FEM formulations is illustrated, e.g., in~Ref.~\cite{WuCangellaris2004MOR}. MOR leverages the observation that the network matrices of resonant structures are predominantly governed by a small set of poles and zeros of the underlying complex-valued rational function. The number of degrees of freedom $N_\mathrm{rd}$ describing the small set of poles and zeros is by far much smaller than the number of degrees of freedom $N$ involved in the~impedance matrix (\ref{eq:Zmat}) or the scattering matrix Eq.~(\ref{eq:Smat}), i.e.\ $N_\mathrm{rd} \ll N$. Since MOR is so important and useful for determining network matrices of resonant structures, the general idea is briefly sketched hereinafter.

For this purpose, the frequency-domain representation of the introduced impedance formulation Eqs.~(\ref{eq:disccurlcurl}) and (\ref{eq:output}) is employed as the starting point. The state vector is expressed by a reduced-order state vector $\underline{\mathbf{x}}_\mathrm{rd}\in\mathbb{C}^{N_\mathrm{rd}}$ using the projection
\begin{equation}
\underline{\mathbf{x}} =\underline{\mathbf{W}}\,\underline{\mathbf{x}}_\mathrm{rd}\textnormal{.}\label{eq:projection}
\end{equation}
The projection matrix $\underline{\mathbf{W}}\in\mathbb{C}^{N\times N_\mathrm{rd}}$ has many more rows than columns and is often chosen to be unitary, i.e.\ $\underline{\mathbf{W}}^\mathrm{H}\,\underline{\mathbf{W}} = \mathbf{I}\in\mathbb{R}^{N_\mathrm{rd}\times N_\mathrm{rd}}$. The superscript $\mathrm{H}$ denotes the complex-conjugate of the matrix. Various ways exist to determine $\underline{\mathbf{W}}$. One prominent approach is to construct an orthonormal basis for the field distributions, which are determined on a finite yet sufficient number of frequency samples:
\begin{equation}
\underline{\mathbf{W}} =\mathrm{orth}(\underline{\mathbf{x}}_1,\underline{\mathbf{x}}_2,\ldots,\underline{\mathbf{x}}_i,\ldots,\underline{\mathbf{x}}_I)\textnormal{.}
\end{equation}
Here, 
$\mathrm{orth}(\quad)$ provides an orthonormal basis and $\underline{\mathbf{x}}_i$ solves Eqs.~(\ref{eq:lse}) and (\ref{eq:inverse}) for the angular frequency $\omega_i$. Sometimes the vectors $\underline{\mathbf{x}}_i$ are also referred to as snapshots of the system. The sampling frequencies are selected to lie within the interval bounded by the minimum and maximum angular frequencies under consideration. i.e.\ $\omega_\mathrm{min}\leq \omega_i\leq\omega_\mathrm{max}$. The primary computational effort of the MOR procedure lies in the computation of the snapshots $\underline{\mathbf{x}}_i$ (full field solutions) and the construction of the projection matrix $\underline{\mathbf{W}}$. Multiplying Eq.~(\ref{eq:Zstatefd}) from the left-hand side by $\underline{\mathbf{W}}^{\mathrm{H}}$ and exploiting the projection Eq.~(\ref{eq:projection}) delivers the reduced-order system
\begin{equation}
j\omega\underbrace{{\underline{\mathbf{W}}^{\mathrm{H}}} \mathbf{M}{\underline{\mathbf{W}}}}_{\underline{\mathbf{M}}_\mathrm{rd}}\,\underline{\mathbf{x}}_\mathrm{rd}
=
\underbrace{\underline{\mathbf{W}}^{\mathrm{H}} \mathbf{A}\underline{\mathbf{W}}}_{\underline{\mathbf{A}}_{\mathrm{rd}}}\,\underline{\mathbf{x}}_\mathrm{rd}+
\underbrace{\underline{\mathbf{W}}^{\mathrm{H}}{\mathbf{B}}}_{\underline{\mathbf{B}}_{\mathrm{rd}}}\,\underline{\mathbf{i}}\textnormal{.}\label{eq:morstate}
\end{equation}
The corresponding output equation reads as
\begin{equation}
\underline{\mathbf{v}} = 
\underbrace{\mathbf{C}\mathbf{W}}_{\underline{\mathbf{C}}_{\mathrm{rd}}}\,\underline{\mathbf{x}}_\mathrm{rd}\label{eq:moroutput}
\textnormal{.}
\end{equation}
The matrices in the reduced system $\underline{\mathbf{M}}_\mathrm{rd}\in\mathbb{C}^{ N_\mathrm{rd}\times N_\mathrm{rd}}$, $\underline{\mathbf{A}}_\mathrm{rd}\in\mathbb{C}^{ N_\mathrm{rd}\times N_\mathrm{rd}}$, $\underline{\mathbf{B}}_\mathrm{rd}\in\mathbb{C}^{ N_\mathrm{rd}\times N_\mathrm{t}}$, $\underline{\mathbf{C}}_\mathrm{rd}\in\mathbb{C}^{N_\mathrm{t}\times N_\mathrm{rd}}$ are significantly smaller than those of the original system. The number of degrees of freedom of the reduced-order model, typically $N_\mathrm{rd}\approx10^2$, depends on the frequency interval $\omega_\mathrm{min}$ to $\omega_\mathrm{max}$ over which the reduced-order model is required to accurately capture the behavior of the original system. Sorting Eq.~(\ref{eq:morstate}) for the reduced-order vector yields
\begin{equation}
\underline{\mathbf{x}}_\mathrm{rd}
=
\left[j\omega\underline{\mathbf{M}}_\mathrm{rd}
-\underline{\mathbf{A}}_\mathrm{rd}\right]^{-1}\underline{\mathbf{B}}_\mathrm{rd}\,\underline{\mathbf{i}}\textnormal{.}
\end{equation}
This vector is readily available on account of the comparably small dimensions and a good approximation of the field distribution is available via Eq.~(\ref{eq:projection}) within the interval from $\omega_\mathrm{min}$ to $\omega_\mathrm{max}$. Combining this expression with Eq.~(\ref{eq:moroutput}) gives
\begin{equation}
\underline{\mathbf{v}}=
\underbrace{\underline{\mathbf{C}}_{\mathrm{rd}}\left[j\omega\underline{\mathbf{M}}_\mathrm{rd}
-\underline{\mathbf{A}}_\mathrm{rd}\right]^{-1}\underline{\mathbf{B}}_\mathrm{rd}}_{\underline{\mathbf{Z}}(j\omega)}\,\underline{\mathbf{i}}\textnormal{.}
\end{equation}
The impedance matrix based on the reduced-order model allows for fast sweeps on the relevant interval with a high frequency resolution. Once the impedance matrices have been determined, they can be converted into the scattering matrices using standard conversion relations (see, e.g., Chapter~4 in~Ref.~\cite{PozarMicrowave}). This allows for accurately sample peaks of resonances with high quality factors. Note that MOR schemes can alternatively be directly applied to the scattering formulation Eqs.~(\ref{eq:states}) and (\ref{eq:outputs}). MOR is also well suited for determining the network matrices of non-resonant RF structures, such as traveling-wave structures.
\newpage
\section{Error Sources in Electromagnetic Simulations\label{sec:errors}}
The following subsections distinguish between model errors and errors in solving the model equations. If possible, the sources of error are illustrated using an example from the field of RF for particle accelerators.

\subsection{Modeling (or Model) Errors}
Modeling errors refer to inaccuracies introduced during the formulation of a mathematical representation of a physical system. They typically arise from idealized material properties, simplified geometries, or the neglect of secondary physical effects. As a result, the governing partial differential equations with their respective boundary and excitation conditions may only approximate the actual system behavior. Examples of modeling errors in modeling of superconducting cavities include neglecting manufacturing errors or non-uniform shrinkage during cooling. Assuming the superconducting niobium sheets to be perfectly electrically conducting is an additional example of an, albeit minor, modeling error.

\subsection{Numerical Errors}
Numerical errors are additional error sources and arise from approximating solutions of model equations. These errors that result when solving the (partial differential) equations can be categorized as follows.

\subsubsection{Geometry Error}
Geometry errors occur when the geometries in the domain are only approximated by the mesh, rather than being represented exactly. In other words, geometry errors arise from the approximation of curved or complicated geometrical features using a hexahedral or tetrahedral mesh. Geometrical errors result from the discrepancy between the exact geometry and material distribution and their approximation in the discrete formulation. In Figure~\ref{fig:tet_mesh}, the smooth, elliptically shaped boundary of the cavity is approximated by a sequence of straight-line segments, while in Fig.~\ref{fig:hex_mesh} a staircase approximation is employed; it is precisely this geometric approximation that introduces the geometrical error. The geometrical error can be reduced by refining the mesh. Moreover, partially filled hexahedral cells or the curved tetrahedral elements are used for an improved description of the smooth boundary. Note that the discretization shown in Fig.~\ref{fig:traingulated_wg} does not introduce a geometry error at all, since the waveguide boundary is exactly represented by (or conforms to) the sequence of straight-line segments of the tetrahedral mesh cells.

\subsubsection{Discretization Error}
The discretization error arises when partial differential equations, defined on a continuous spatial domain, are approximated on a finite mesh or grid, as in the transition from Eqs.~(\ref{eq:contcurlcurl}) to (\ref{eq:disccurlcurl}). The continuous electromagnetic fields are described by basis functions with a finite set of degrees of freedom or by sampling field components at the midpoints of edges, faces, or volumetric elements of the mesh. Additionally, the~spatial derivatives are approximated on this discrete representation of the electromagnetic fields. The~discretization requires a sufficiently large number of degrees of freedom, allowing this quasi-continuous approximation to capture the properties of the continuous field solution. The discretization error depends on the mesh size, the element type, and the order of the basis functions. The discretization error typically decreases (and computational burden increases) as the discretization is refined or higher-order schemes are used. For a sufficiently refined discretization, the properties of the discrete field solution should become independent of the mesh, reflecting the behavior of the underlying continuous problem, where no mesh exists at all. The discretization should employ enough degrees of freedom to achieve a reasonable approximation of the field solution, while keeping their number as low as possible to minimize computational costs. In practice, this balance is usually achieved via iterative mesh refinement. The mesh is refined in a stepwise manner until the field solution converges to a mesh-independent approximation.
 
\subsubsection{Round-off Error}
Round-off errors arise because digital computers are inherently limited to a finite set of states. Real (and imaginary) numbers are stored in a dedicated floating-point format with a finite number of bits. Thus, the~numbers cannot be represented exactly in digital form and arithmetic operations must be performed approximately. Round-off errors can often be neglected when the problem under study is well-conditioned. In contrast, round-off errors are a major challenge for ill-conditioned problems, including systems of linear equations, whose matrices are characterized by large condition numbers [see Eq.~(\ref{eq:inequality}) in Section~\ref{sec:freq_do}].

\section{Remarks on the Practice of Electromagnetic Simulations\label{sec:remarkspractice}}
Complementary to the theoretical considerations outlined above, few important remarks regarding the practical execution of electromagnetic simulations are provided.

\subsection{Setting up a Simulation for a New Type of Problem}
Setting up simulations for unfamiliar problems can be difficult. Nowadays commercial software packages, such as CST Studio Suite\textregistered~\cite{CST}, are very flexible and support various methods, various meshes, such as hexahedral or tetrahedral, and time-domain and frequency-domain methods. Many options are associated with each solver and mesh type, as well as with frequency- and time-domain methods, and default settings do not necessarily yield optimal results. The meaning and effect of these options can be unclear since commercial software is essentially a "black box" for users and help files are not always informative. It is therefore recommended to begin the analysis with a simplified version of the problem. For instance, losses may initially be neglected. Furthermore, the geometry may be idealized by omitting certain features to obtain a symmetric structure. In such cases, symmetry planes can be exploited to reduce the computational domain, or the problem can be simplified to a two-dimensional or even one-dimensional formulation. It is often advisable to start with a coarse discretization to assess the demands of the problem in terms of computational resources and anticipated runtime. The simplified problem comparatively quickly allows for the evaluation of different solvers and their respective configurations. Once an appropriate solver and a suitable configuration are selected, the problem in its full complexity (with a reasonable mesh) can be addressed. The balance between model accuracy/complexity and available hardware resources is sometimes challenging to find, in particular for large problems. Therefore, various attempts are typically required to determine an appropriate simulation setup for a given problem. However, how to decide whether the delivered solution is reasonable? The solution is a priori unknown. If it were known, there would be no need for simulation.

\subsection{Validation and Critical Assessment of the Results\label{sec:validation}}
As the computed field distributions and derived quantities are affected by numerical errors as outlined in Section~\ref{sec:errors}, their accuracy requires validation. Validation is typically performed by comparing the simulation results with analytical estimations, experimental measurements, or with results obtained from other numerical methods. In addition, studies, such as mesh refinement, reduction of the stopping criteria for the time-marching scheme, or reduction of the relative residual, are often carried out to confirm that the~numerical solution is valid.

The following list outlines several important points and observations on the validation and on critical assessment of the results:
\begin{itemize}
 \item \underline{Never ever blindly believe in the solutions provided by numerical codes}, since inappropriate assumptions, and/or inappropriate methods and/or wrong settings for the problem under study can lead to unreasonable results.
 \item Leverage as much theoretical knowledge related to the problem as possible. Although closed-form analytical solutions are often unavailable, theoretical insight is invaluable for assessing the validity of results (e.g., conservation of energy and charges, reciprocity, causality, etc.).
 \item Compare models of varying accuracy, ranging from simple analytical estimations to numerical schemes with different settings, to assess the consistency of their results.
 \item Always conduct mesh refinement studies to verify that simulation results are independent of the~discretization. Ensure that the mesh adequately represents the structure and that the smallest wavelength is resolved by a sufficient number of degrees of freedom (at least $10$ nodes when using FIT on a hexahedral grid).
 \item Always read the solver log file! It provides valuable information! Carefully account for warnings, not just for errors! This is particularly relevant, as the threshold for commercial solvers to report an error can be very high.
 \item When waveguide ports are used to excite a domain with rectangular, circular, or coaxial waveguides, the field patterns of the waveguide modes, wave impedances, cutoff frequencies and propagation constants should be compared with analytical formulas. If the excitation of the computational domain is not accurate, the simulation outcomes will not be accurate either.
  \item Use all possibilities to compare simulations with measurements. Future simulations will benefit from these comparisons. Neither measurements nor simulations are right or wrong and both can be subject to systematic errors. If it is not possible to achieve a reasonable agreement between measurement and simulations, the problem is not well understood. Carefully evaluate how both simulations and measurements can be improved.
\end{itemize}

\section{Conclusion\label{sec:conclusions}}
This contribution reviews key concepts of electromagnetic simulations for RF applications in particle accelerators. While detailed derivations of specific methods are not provided, general procedures for conducting simulations are outlined from a theoretical and practical point of view. Challenges for simulating highly resonant structures are discussed and model-order reduction techniques are sketched to address these challenges. Common sources of error occurring in electromagnetic simulations are listed and explained. Important remarks on the practice of electromagnetic simulations are also provided. In this manner, both conceptual understanding and practical insights are conveyed. The attached list of references aims to serve as a valuable resource for readers seeking further literature related to CEM. 

\section*{Acknowledgement}
The author would like to kindly thank the organizers of the CERN accelerator school for the opportunity to give a presentation and to prepare this report. Moreover, the author would like to express his appreciation to Sosoho-Abasi Udongwo from Brandenburg University of Technology, Kai Papke from DESY, and Shahnam Gorgi Zadeh from CERN for a "friendly review" of this this document.

Sosoho-Abasi Udongwo kindly contributed Fig.~\ref{fig:freqdo_example}. The cavity model with couplers in Fig.~\ref{fig:network} was provided by N.~Eddy from Fermi National Accelerator Laboratory. The cavity model with couplers in Fig.~\ref{fig:wake} was provided by A.\ V\'elez Saiz from Helmholtz-Zentrum Berlin. The author acknowledges the use of OpenAI's ChatGPT and DeepL, which were employed to assist with language editing and to improve the clarity, readability, and presentation of the text.


\newpage
\begin{thebibliography}{99}

\bibitem{rylander2012computational} T. Rylander, P. Ingelström, and A. Bondeson, Computational Electromagnetics, Texts in Applied Mathematics, vol.~51, Springer Science+Business Media, New York, 2nd~ed., 2012, doi: 10.1007/978-1-4614-5351-2.

\bibitem{sheng2012essentials} X.‑Q. Sheng and W. Song, Essentials of Computational Electromagnetics, Wiley‑IEEE Press, Hoboken, NJ, USA, 2012, doi: 10.1002/9780470829646.

\bibitem{Davidson2005} D. B. Davidson, Computational Electromagnetics for RF and Microwave Engineering, Cambridge University Press, Cambridge, 2005.

\bibitem{Sumithra2017CEMReview} P.~Sumithra and D.~Thiripurasundari, Review on Computational Electromagnetics, Advanced Electromagnetics, vol.~6, no.~3, pp.~1--10, 2017, doi: 10.7716/aem.v6i3.407.

\bibitem{Weiland2017} T. Weiland, Electromagnetic simulators -- status and future directions, IET Science, Measurement and Technology, vol. 11, no. 6, pp. 681--686, 2017, doi: 10.1049/iet-smt.2017.0085.

\bibitem{Munteanu2010}
I. Munteanu, M. B. Timm, and T. Weiland, It's about time, IEEE Microwave Magazine, vol. 11, no. 2, pp. 60--69, May 2010, doi: 10.1109/MMM.2010.935775. 

\bibitem{Weiland2008} T. Weiland, M. Timm, and I. Munteanu, A practical guide to 3-D simulation, IEEE Microwave Magazine, vol. 9, no. 6, pp. 62--75, Dec. 2008, doi: 10.1109/MMM.2008.929772.

\bibitem{yee1966numerical} K. S. Yee, Numerical solution of initial boundary value problems involving Maxwell's equations in isotropic media, IEEE Transactions on Antennas and Propagation, vol. 14, no. 3, pp. 302--307, 1966, doi: 10.1109/TAP.1966.1138693.

\bibitem{taflove2000computational} A. Taflove and S. C. Hagness, Computational Electrodynamics: The Finite-Difference Time-Domain Method, Artech House, Norwood, MA, USA, 2nd~ed., 2000.

\bibitem{SchuhmannWeiland2004} R. Schuhmann and T. Weiland, Recent advances in finite integration technique for high frequency applications, in Scientific Computing in Electrical Engineering, W. H. A. Schilders, E. J. W. ter Maten, and S. H. M. J. Houben, Eds., Springer Berlin Heidelberg, Mathematics in Industry, vol. 4, pp. 46--57, 2004, doi: 10.1007/978-3-642-55872-6-4.

\bibitem{weiland1996timedomain} T. Weiland, Time domain electromagnetic field computation with finite difference methods, Int. J. Numer. Model., vol. 9, no. 4, pp. 295--319, 1996.

\bibitem{weiland1977discretization} T. Weiland, A discretization method for the solution of Maxwell's equations for six-component fields, Electron. Commun. (AEÜ), vol. 31, no. 3, pp. 116--120, 1977.

\bibitem{Jin} J. M. Jin, The Finite Element Method in Electromagnetics, 3rd ed. Hoboken, NJ: Wiley, 2014, doi: 10.5555/2655347.

\bibitem{volakis1998electromagnetics} J. L. Volakis, Electromagnetics Through the Finite Element Method, John Wiley \& Sons, New York, 1998.

\bibitem{polycarpou2006introduction} A. C. Polycarpou, Introduction to the Finite Element Method in Electromagnetics, Morgan \& Claypool Publishers, 2006.

\bibitem{gibson2008method} W. C. Gibson, The Method of Moments in Electromagnetics, CRC Press, 2008.

\bibitem{chew2001fast} W. C. Chew, J.-M. Jin, E. Michielssen, and J. Song, Fast and Efficient Algorithms in Computational Electromagnetics, Artech House, Boston, MA, 2001.

\bibitem{CST} Dassault Systemes, CST Studio Suite,\\available at \url{https://www.3ds.com/products/simulia/cst-studio-suite}.

\bibitem{HFSS} ANSYS Inc., Ansys HFSS,\\available at \url{https://www.ansys.com/products/electronics/ansys-hfss}.

\bibitem{COMSOL} COMSOL Inc., COMSOL Multiphysics,\\available at \url{https://www.comsol.com/comsol-multiphysics}.

\bibitem{EMPIRE} IMST GmbH, EMPIRE XPU, available at \url{https://empire.de/}.

\bibitem{EMPro} Keysight Technologies, Keysight EMPro, available at \url{https://www.keysight.com/}.

\bibitem{openEMS} Thorsten Liebig, openEMS -- Open Electromagnetic Field Solver, available at \newline \url{https://www.openems.de}.

\bibitem{openCFS} openCFS Association, openCFS -- Coupled Field Simulation, available at \newline \url{https://www.opencfs.org}.

\bibitem{Meep} A. F. Oskooi et al., Meep -- MIT Electromagnetic Equation Propagation, available at \newline \url{https://meep.readthedocs.io}.

\bibitem{FEniCS} The FEniCS Project, FEniCS -- Automated Solution of Differential Equations, available at \newline \url{https://fenicsproject.org}.

\bibitem{NGSolve} J. Sch\"oberl, Netgen/NGSolve, available at \url{https://ngsolve.org}.

\bibitem{ZhangLi2008} K.~Zhang and D.~Li, Electromagnetic Theory for Microwaves and Optoelectronics, 2nd~ed., Springer, Berlin, 2008, doi:10.1007/978-3-540-74296-8.

\bibitem{shreyber2024emtheory} I.~Shreyber, Electromagnetic Theory, Proc.\ CAS–CERN Accelerator School: Introduction to Accelerator Physics, M.~Filippova, D.~Rivoiron, H.~Schmickler, and F.~Tecker, Eds., CERN Accelerator School, Santa Susanna, Spain, 22\,Sep--5\,Oct\,2024, CERN, 2024, doi:10.17181/CAS2024SantaSusanna.

\bibitem{flisgen2018recap} T.~Flisgen, J.~Heller, and U.~van~Rienen, Recapitulation of electromagnetism, Proc.\ CAS–CERN Accelerator School: Beam Injection, Extraction and Transfer, B.~Holzer, Ed., CERN Yellow Reports: School Proceedings, Vol.~5/2018, Erice, Italy, 10--19 March 2017, pp.\ 69--88, CERN‑2018‑008‑SP, doi:10.23730/CYRSP‑2018‑005.69.

\bibitem{wolski2011emfields} A.~Wolski, Theory of electromagnetic fields, Proc.\ CAS–CERN Accelerator School: RF for Accelerators , R.~Bailey, Ed. CERN Yellow Reports: School Proceedings, Geneva, 2011, pp.\ 15--65, presented at CAS 2010, Ebeltoft, Denmark, 8--17 June 2010, doi:10.5170/CERN-2011-007.15. 

\bibitem{miles2000rf} J.~Miles, Proc.\ CAS - CERN Accelerator School: Radio Frequency Engineering: Proceedings, Seeheim, Germany, 8 -- 16 May 2000, CERN Accelerator School Course on RF Engineering, Seeheim, Germany, 2000, CERN 2005-003, doi: 10.5170/CERN-2005-003.

\bibitem{Griffiths} D. J. Griffiths, \textit{Introduction to Electrodynamics}, Prentice-Hall International (UK) Limited, London, 1999.

\bibitem{Jackson} J. D. Jackson, \textit{Classical Electrodynamics}, Wiley, New York 1998.

\bibitem{Collin2} R. E. Collin, \textit{Foundations for Microwave Engineering}, McGraw-Hill Ryerson Limited, 1992.

\bibitem{Collin1} R. E. Collin, \textit{Field Theory of Guided Waves}, Second Edition, IEEE Press, New York, USA, 1991.

\bibitem{Balanis} C. A. Balanis, \textit{Advanced Engineering Electromagnetics}, John Wiley \& Sons, New York, USA, 1989.

\bibitem{SteinmetzKurzClemens2011} T.~Steinmetz, S.~Kurz, and M.~Clemens, Domains of validity of quasistatic and quasistationary field approximations, COMPEL: The International Journal for Computation and Mathematics in Electrical and Electronic Engineering, vol.~30, no.~4, pp.~1237--1247, 2011, doi:10.1108/03321641111133154.

\bibitem{Vogel} E. Vogel et al., Considerations on the third harmonic RF of the European XFEL, Proceedings of the 13th International Workshop on RF Superconductivity 2007, Beijing, China, 2007, pp. 481--485.

\bibitem{DeyMittra1997} S.~Dey and R.~Mittra, A locally conformal finite-difference time-domain (FDTD) algorithm for modeling three-dimensional perfectly conducting objects, IEEE Microwave and Guided Wave Letters, vol.~7, no.~9, pp.~273--275, 1997, doi:10.1109/75.640131.

\bibitem{ZagorodnovSchuhmannWeiland2003} I.~A.~Zagorodnov, R.~Schuhmann, and T.~Weiland, A uniformly stable conformal FDTD‑method on Cartesian grids, International Journal of Numerical Modelling: Electronic Networks, Devices and Fields, vol.~16, no.~2, pp.~127--141, 2003, doi:10.1002/jnm.488.

\bibitem{SchreiberClemensVanRienen2004} U.~Schreiber, M.~Clemens, and U.~van Rienen, Conformal FIT formulation for simulations of electro‑quasistatic fields, International Journal of Applied Electromagnetics and Mechanics, vol.~19, no.~1‑4, pp.~193--197, 2004, doi:10.3233/JAE-2004-561.

\bibitem{Wittig1} T. Wittig, I. Munteanu, R. Schuhmann, and T. Weiland, Two-step Lanczos algorithm for model order reduction, in IEEE Transactions on Magnetics, vol. 38, no. 2, pp. 673--676, March 2002, doi: 10.1109/20.996175.

\bibitem{Wittig2} T. Wittig, R. Schuhmann, and T. Weiland, Model order reduction for large systems in computational electromagnetics, Linear Algebra and its Applications, Volume 415, Issues 2--3, pp. 499-530, 2006, pp. 499-530, doi: 10.1016/j.laa.2004.06.023.

\bibitem{vanRienen1999} U. van Rienen, Finite integration technique on triangular grids revisited, International Journal of Numerical Modelling: Electronic Networks, Devices and Fields, vol. 12, pp. 107--128, 1999.

\bibitem{Flisgen_TMMT} T. Flisgen, H. -W. Glock and U. van Rienen, Compact Time-Domain Models of Complex RF Structures Based on the Real Eigenmodes of Segments, in IEEE Transactions on Microwave Theory and Techniques, vol. 61, no. 6, pp. 2282--2294, June 2013, doi: 10.1109/TMTT.2013.2260765.

\bibitem{Klingbeil} H. Klingbeil, Ferrite cavities, CAS - CERN Accelerator School: RF for Accelerators, pp.~299--317, 2010, doi: 10.5170/CERN-2011-007.29.

\bibitem{GriffithsHigham2010} D. F. Griffiths and D. J. Higham, Numerical Methods for Ordinary Differential Equations: Initial Value Problems, 2nd ed., Springer, London, 2010.

\bibitem{HairerNorsettWanner1993} E. Hairer, S. P. N{\o}rsett, and G. Wanner, Solving Ordinary Differential Equations I: Nonstiff Problems, 2nd ed., Springer, Berlin, 1993.

\bibitem{Papke2017} K.~Papke, F.~Gerigk, and U.~van~Rienen, Comparison of coaxial higher order mode couplers for the CERN Superconducting Proton Linac study, Phys. Rev. Accel. Beams, vol.~20, p.~060401, 2017.

\bibitem{bathe1996finite} K. J. Bathe, Finite Element Procedures, Prentice Hall, Upper Saddle River, NJ, USA, 1996.

\bibitem{egger2021second} H. Egger and B. Radu, A second‑order finite element method with mass lumping for Maxwell's equations on tetrahedra, SIAM Journal on Numerical Analysis, vol.~59, no.~2, pp.~864--885, 2021, doi:10.1137/20M1318912.

\bibitem{schnepp2012efficient} S. M. Schnepp and T. Weiland, Efficient large scale electromagnetic simulations using dynamically adapted meshes with the discontinuous Galerkin method, Journal of Computational and Applied Mathematics, vol. 236, no. 18, pp. 4909--4924, 2012, doi: 10.1016/j.cam.2011.12.005.

\bibitem{hesthaven2008nodal} J. S. Hesthaven and Tim Warburton, Nodal Discontinuous Galerkin Methods: Algorithms, Analysis, and Applications, Texts in Applied Mathematics, vol. 54, Springer Science+Business Media, New York, NY, USA, 2008, doi: 10.1007/978-0-387-72067-8.

\bibitem{phung2019microwave} G. N. Phung, F. J. Schmückle, R. Doerner, B. Kähne, T. Fritzsch, U. Arz, and W. Heinrich,
\newblock Influence of microwave probes on calibrated on‑wafer measurements, IEEE Transactions on Microwave Theory and Techniques, vol. 67, no. 5, pp. 1892--1900, May 2019, doi:10.1109/TMTT.2019.2903400.

\bibitem{rahimof2024study} Y. Rahimof, I. A. Nechepurenko, M. R. Mahani, A. Tsarapkin, and A. Wicht, The study of 3D FDTD modelling of large-scale Bragg gratings validated by experimental measurements, Journal of Physics: Photonics, vol.~6, no.~4, p.~045024, 2024, doi:10.1088/2515-7647/ad8824.

\bibitem{meister2026numerical} A. Meister, Numerical Methods for Linear Systems of Equations: An Introduction to Modern Methods with MATLAB Implementations by C. Voemel, Springer Fachmedien Wiesbaden, 2026, doi: 10.1007/978-3-658-50260-7.

\bibitem{FlisgenACC39} T. Flisgen, H.-W. Glock, P. Zhang, I. R. R. Shinton, N. Baboi, R. M. Jones, and U. van Rienen, Scattering parameters of the 3.9 GHz accelerating module in a free-electron laser linac: A rigorous comparison between simulations and measurements, Phys. Rev. ST Accel. Beams 17, 2014, doi: 10.1103/PhysRevSTAB.17.022003.

\bibitem{PozarMicrowave} D. M. Pozar, Microwave Engineering, 4th ed., John Wiley and Sons, 2012.

\bibitem{SchuhmannWeiland2000} R. Schuhmann and T. Weiland, Rigorous analysis of trapped modes in accelerating cavities, Phys. Rev. Special Topics -- Accelerators and Beams, vol. 3, p. 122002, 2000, doi: 10.1103/PhysRevSTAB.3.122002.

\bibitem{Padamsee2017} H. Padamsee, 50 years of success for SRF accelerators -- a review, Superconductor Science and Technology, vol. 30, no. 5, p. 053003, 2017, doi: 10.1088/1361-6668/aa6376.

\bibitem{Baur2014MOR} U. Baur, P. Benner, and L. Feng, Model Order Reduction for Linear and Nonlinear Systems: A System-Theoretic Perspective, Archives of Computational Methods in Engineering, vol. 21, pp. 331--358, 2014, doi: 10.1007/s11831-014-9111-2.

\bibitem{Simeoni2009MOR}
M. Simeoni, G. A. E. Vandenbosch, and I. E. Lager, Model-order reduction techniques for linear electromagnetic problems -- an overview, IEEE Transactions on Magnetics, vol. 45, no. 3, pp. 1400--1403, 2009, doi: 10.1109/TMAG.2009.2012698.

\bibitem{WuCangellaris2004MOR} H. Wu and A. C. Cangellaris, Model-order reduction of finite-element approximations of passive electromagnetic devices including lumped electrical-circuit models, IEEE Transactions on Microwave Theory and Techniques, vol. 52, no. 9, pp. 2305--2313, 2004, doi: 10.1109/TMTT.2004.834582.

\end{thebibliography}
\end{document}